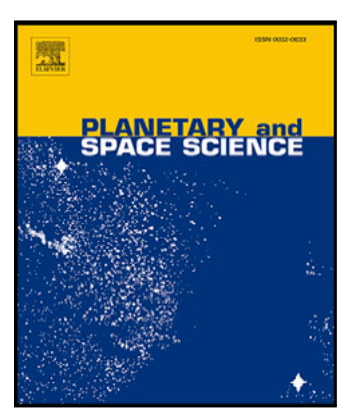

# Revisiting the dynamics of the Prometheus–Pandora system

Demétrio Tadeu Ceccatto [a,b,*], Nelson Callegari Jr. [a], Gabriel Teixeira Guimarães [c,d]

[a] São Paulo State University (UNESP), Institute of Geosciences and Exact Sciences, Av. 24A, 1515, Rio Claro, 13506-900, São Paulo, Brazil
[b] Universidade Virtual do Estado de São Paulo (UNIVESP), Avenida Paulista, 352, 01310-905, São Paulo, Brazil
[c] Graduate University for Advanced Studies (SOKENDAI), Shonankokusaimura, Hayama, Miura District, Kanagawa, 240-0115, Japan
[d] National Astronomical Observatory of Japan (NAOJ), Osawa, Mitaka, Tokyo, 181-8588, Japan

ABSTRACT

This study revisits the dynamics of Prometheus and Pandora. The analysis focuses on their current orbits and considers mutual perturbations and the non-spherical shape of Saturn. Numerical integrations of the full equations of motion were performed in the context of the N-body problem using a dense set of initial conditions. Spectral analysis of these orbits, together with an interpretation of the dynamical maps, revealed several resonance domains of various orders between Prometheus and Pandora, including the 121:118 resonance domain. Resolving the secular equations for eccentricity and inclination revealed the absence of a forced component in both cases. The period associated with the precession rate, $\Delta\varpi$, corresponds to the anti-alignment period. Chaos was quantified by analysing the diffusion of independent frequencies, which indicates dynamics governed by weak chaos. Minor perturbations caused by the Janus–Epimetheus pair on the orbits of Prometheus and Pandora were also considered. Finally, we investigated the dynamical history of the Prometheus–Pandora system, accounting for Pandora's migration due to tidal interactions with Saturn. This revealed several past cross-resonances with a low probability of capture, as well as a scenario in which the Prometheus–Pandora pair exhibited a co-orbital configuration.

## 1. Introduction

Saturn's inner satellite system is characterised by unique dynamic complexity resulting from intense gravitational interactions and mean-motion resonances (MMRs), which determine the stability of its orbits. Although these resonances can lead to significant orbital stability, they can also induce unstable and chaotic behaviour depending on the specific conditions of each system (see, for example, Callegari and Yokoyama, 2010; El Moutamid et al., 2014; Munõz-Gutiérrez and Giuliatti Winter, 2017; Callegari and Yokoyama, 2020; Callegari et al., 2021; Rodríguez and Callegari, 2021; Callegari and Rodríguez, 2023; Ceccatto et al., 2025). One such example is the pair of moons Prometheus and Pandora.

The pair Prometheus (S XVI) and Pandora (S XVII), discovered by the *Voyager* mission in 1980–1981, stand out for their position on either side of the narrow F ring. Initially described as shepherd satellites, subsequent studies revealed that their orbital evolution is governed by much more complex tidal forces and mutual interactions than predicted.

Poulet and Sicardy (2001) furthered the investigation of the long-term evolution of this pair by demonstrating that, under the effect of density torques and tidal dissipation, the orbits converge rapidly. Based on the actual configurations of the system, their findings suggest that the pair has a lifetime of approximately 20 million years before an unavoidable close encounter occurs, unless an orbital resonance capture happens.

The chaotic nature of the system became evident following observations by the *Hubble Space Telescope* (HST) in 1995, which revealed a discrepancy between Prometheus' longitude and *Voyager* predictions. French et al. (2003) confirmed that this deviation was anti-alignment, with Pandora ahead of its orbit, indicating a mutual exchange of energy and angular momentum. Goldreich and Rappaport (2003a,b), identified the physical origin of this chaos as the superposition of four components of the 121:118 resonance. The intensity of these interactions peaks during the anti-alignment of the pericentre, which occurs every 6.2 years.

Subsequent studies have expanded on this understanding: Cooper and Murray (2004) integrated the effects of the eight main satellites, confirming the impact of Mimas' 3:2 resonance on Pandora's orbit, while Renner et al. (2005) refined the masses and densities of the moons. Building on this theory, Farmer and Goldreich (2006) explained

* Corresponding author at: São Paulo State University (UNESP), Institute of Geosciences and Exact Sciences, Av. 24A, 1515, Rio Claro, 13506-900, São Paulo, Brazil.

*E-mail address:* dt.ceccatto@unesp.br (D.T. Ceccatto).

that the "jumps" in the mean-motion are short episodes of libration that occur when the potential amplitude is at its maximum, using the analogy of a parametric pendulum.

Building on this research, the present study revisits the dynamics of the Prometheus–Pandora system by performing numerical integrations of the N-body problem. The study is organised as follows: Section 2 details the methodology adopted, including the use of geometric elements and spectral analysis. Section 3 discusses the analysis of current orbits, focusing on mutual perturbative effects and the influence of the Mimas–Tethys pair. Section 4 presents an investigation of chaotic motion, quantified by calculating the fundamental frequency diffusion coefficient through the wavelet transform. Section 5 addresses the gravitational perturbations exerted by Janus and Epimetheus. Section 6 explores the dynamic past of the system by examining the migration of Pandora under the influence of tides and the potential for an ancient co-orbital configuration. Section 7 finally presents the main conclusions of this work.

## 2. Methodology

The main objective of this study is to analyse the global dynamics of the Prometheus–Pandora system through the construction of dynamical maps in the space of fundamental frequencies. Fig. 1(a) illustrates the resulting map, where the initial conditions $(a_0, e_0)$ are distributed over a regular grid in the plane of eccentricity versus semi-major axis.

For each initial condition, we numerically integrate the full equations of motion considering Saturn as a non-spherical central body, including the zonal harmonics $J_2$ and $J_4$ of its gravitational potential, as well as the mutual gravitational interactions between Prometheus and Pandora. In an extended model, additional perturbations due to Janus, Epimetheus, Mimas, and Tethys can also be included to assess their influence on the system dynamics.

The equations of motion follow the Hamiltonian formulation described in Callegari and Yokoyama (2010), which we summarise here for completeness. We consider $N$ satellites with masses $m_i$ $(i = 1, \dots, N)$ orbiting a central oblate planet of mass $M$. Denoting by $\vec{r}_i$ the position vectors of the satellites relative to the planet's centre and by $\vec{p}_i$ their corresponding momenta relative to the barycentre of the system, the pairs $(\vec{r}_i, \vec{p}_i)$ define canonical variables in the Poincaré coordinate system. The Hamiltonian is written as

$$H = H_0 + H_1 + H_J, \tag{1}$$

where

$$H_0 = \sum_{i=1}^{N} \left( \frac{|\vec{p}_i|^2}{2\beta_i} - \frac{\mu_i\,\beta_i}{|\vec{r}_i|} \right), \tag{2}$$

$$H_1 = \sum_{0<i<j} \left( -\frac{G\,m_i\,m_j}{\Delta_{ij}} + \frac{\vec{p}_i \cdot \vec{p}_j}{M} \right), \tag{3}$$

$$H_J = \sum_{i=1}^{N} \frac{m_i\,\beta_i}{|\vec{r}_i|} \sum_{l=2}^{4} J_l \left( \frac{R_S}{|\vec{r}_i|} \right)^l P_l(\sin\phi_i), \tag{4}$$

with $\beta_i = M\,m_i/(M + m_i)$, $\mu_i = G(M + m_i)$, and $\Delta_{ij} = |\vec{r}_i - \vec{r}_j|$. Here, $G$ is the gravitational constant, $J_l$ are the zonal coefficients of Saturn, $R_S$ is its equatorial radius, $P_l$ are the Legendre polynomials, and $\phi_i$ are the latitudes of the satellites with respect to Saturn's equator, which is adopted as the reference plane. The term $H_0$ describes the Keplerian motion, $H_1$ accounts for mutual interactions, and $H_J$ represents the perturbation due to the planet's oblateness.

The equations of motion are obtained from Hamilton's canonical equations

$$\frac{d\vec{r}_i}{dt} = \frac{\partial H}{\partial \vec{p}_i}, \qquad \frac{d\vec{p}_i}{dt} = -\frac{\partial H}{\partial \vec{r}_i}, \qquad i = 1, \dots, N, \tag{5}$$

and are numerically integrated using the RA15 integrator (Everhart, 1985). Integrations are performed over a timespan of $T = 516.75$ years with an output step of 0.18 days.

**Table 1**
Physical constants for Saturn.

| Constant | Numerical value |
|---|---|
| $GM$[a] (km$^3$ s$^{-2}$) | $3.7931208 \times 10^7$ |
| Equatorial radius[a] (km) | 60, 268 |
| $J_2$[b] | $16,290.71 \times 10^{-6}$ |
| $J_4$[b] | $-935.83 \times 10^{-6}$ |
| $J_6$[b] | $86.14 \times 10^{-6}$ |

[a] *Horizons* system.
[b] (Jacobson et al., 2006).

To characterise the dynamical behaviour, we construct a spectral dynamical map using the Fast Fourier Transform (FFT) method (Press et al., 1988). For each orbit, a spectral analysis is performed on a selected *variable of analysis* (VA), defined here as semi-major axis, eccentricity, or orbital inclination. The resulting power spectrum is used to compute the spectral number **N**, defined as the number of peaks with amplitudes above a reference amplitude (RA).

The RA is fixed at 5%, and only frequencies corresponding to periods longer than 15 days are retained to filter out short-period oscillations associated with fast orbital terms. An upper cutoff **N*** is introduced such that all values above this threshold are represented by the same colour scale. In the resulting map, low values of **N** (associated with regular motion) appear in lighter tones, whereas high values (indicative of chaotic behaviour) appear darker. This representation allows for the identification of resonant structures and chaotic regions in phase space. More details on this technique can be found in Callegari and Yokoyama (2020), Callegari et al. (2021), Callegari and Rodríguez (2023).

The MERCURY package (Chambers, 1999) is used to compute the osculating orbital elements, namely the semi-major axis, eccentricity, inclination, argument of pericentre, longitude of the ascending node, and mean anomaly. Since Saturn's oblateness induces short-period variations in these elements, we convert them into geometric elements following the procedure described in Renner and Sicardy (2006). This is done by first reconstructing the state vector[1] and then applying the transformation to remove short-period contributions associated with the zonal harmonics.

Using geometric elements provides a more stable representation that is less susceptible to fluctuations. This enables a more precise analysis of long-term orbital dynamics. Geometric elements also make it easier to identify dynamic patterns that may be obscured in osculating representations. This approach improves our understanding of the interactions within the Prometheus–Pandora system and their impact on satellite orbital dynamics.

Table 1 shows the values of the physical constants of Saturn used in this work. The initial conditions and masses of the satellites Prometheus, Pandora, Janus, Epimetheus, Mimas, and Tethys are presented in Table 2. The initial conditions were obtained from the *Horizons* ephemeris system (http://ssd.jpl.nasa.gov/horizons.cgi). The mass and mean radius data were adopted from Thomas and Helfenstein (2020), except for Mimas and Tethys, whose values were also derived from the *Horizons* system.

## 3. Dynamical map: the orbital neighbourhood of Prometheus and Pandora

Fig. 1 shows the dynamical map for the orbital neighbourhood of Prometheus (a,c) and Pandora (b,d). The current orbits of Prometheus and Pandora are represented by the red and black stars, respectively.

[1] This is the vector containing the position coordinates and respective instantaneous velocities of the satellite or particle, obtained from the osculating elements of the numerical simulation described above.

**Table 2**
Osculating elements and orbital period for Prometheus, Pandora, Janus, Epimetheus, Mimas, and Tethys, as obtained from the *Horizons* ephemeris system on 1 January 2000 (accessed on 20 October 2024).

| | Symbol | Prometheus | Pandora | Janus | Epimetheus | Mimas | Tethys |
|---|---|---|---|---|---|---|---|
| *Mass* (kg) | $m$ | $1.56 \times 10^{17}$ | $1.37 \times 10^{17}$ | $1.90 \times 10^{20}$ | $5.26 \times 10^{18}$ | $3.75 \times 10^{19}$ | $6.176 \times 10^{20}$ |
| *Mean radius* (km) | $R_m$ | 42.8 | 40.0 | 89.0 | 58.6 | 198.8 | 536.3 |
| *Orbital period* (degree/day) | $P$ | 0.613 | 0.629 | 0.695 | 0.695 | 0.940 | 1.888 |
| *Semi-major axis* (km) | $a$ | 140,024.64 | 142,346.13 | 152,031.24 | 152,089.00 | 186,009.56 | 294,975.42 |
| *Eccentricity* | $e$ | 0.0025 | 0.0015 | 0.0043 | 0.0119 | 0.0176 | 0.0008 |
| *Orbital inclination* (degree) | $i$ | 0.0042 | 0.050 | 0.0166 | 0.3546 | 1.571 | 1.094 |
| *Argument of pericentre* (degree) | $\omega$ | 201.36 | 332.16 | 339.93 | 64.99 | 327.90 | 186.52 |
| *Longitude of the ascending node* (degree) | $\Omega$ | 309.14 | 149.99 | 155.58 | 193.93 | 173.47 | 259.90 |
| *Mean anomaly* (degree) | $\ell$ | 115.18 | 288.84 | 249.21 | 311.04 | 215.95 | 5.29 |

When Pandora is considered the main disturbing body, Fig. 1(a) reveals several dark diagonal structures associated with high **N** values, within which lighter regions corresponding to low **N** values can be identified. These patterns are associated with resonances of different orders between the Prometheus–Pandora pair.

By selecting initial conditions ($a_0$, $e_0$) located in the lighter regions indicated by coloured circles in Fig. 1(a), we identify three mean-motion resonances: 39:38, 40:39, and 41:40. We also identify the domains associated with the Lindblad (**L1**, **L2**) and corotation (**C1**, **C3**) resonances by analysing the respective geometric angles:

(i) **Lindblad angle**: $\phi_1^{m+1:m} = (m+1)\lambda_{\text{Pand}} - m\lambda_{\text{cPro}} - \varpi_{\text{cPro}}$,
(ii) **Corotation angle**: $\phi_2^{m+1:m} = (m+1)\lambda_{\text{Pand}} - m\lambda_{\text{cPro}} - \varpi_{\text{Pand}}$,

where $\lambda$ represents the mean longitude, $\varpi$ is the pericentre argument, and the subscripts $Pand$ and $cPro$ refer to Pandora and a clone of Prometheus, respectively, $m = 38, 39, 40$.

The time variation for these angles is shown in Fig. 2. Figs. 2(a) and (c) show the oscillation of the Lindblad angle around 0, while Figs. 2(b) and (d) show the oscillation of the corotation angle around 180°. Additionally, two second-order resonances, 79:77 and 83:81, have been identified in Fig. 1(a).

Fig. 1(b) shows the orbital neighbourhood of Pandora, assuming that Prometheus is the only primary perturbing body. Similarly to Prometheus, Pandora's orbital region exhibits several dark diagonal structures, two of which encompass distinct lighter patches. A detailed analysis of these bright regions reveals the dynamical domains associated with two second-order resonances: 80:78 and 82:80.

Prometheus's current orbit lies in a region characterised by a high **N** value, corresponding to the separatrices of the 121:118 resonance with Pandora. Fig. 3(a) provides a detailed view of this region, as previously shown in Section 1(a), allowing closer examination of the local resonant dynamics and the surrounding phase-space structures. Prometheus' position on this map allows us to quantitatively explain the irregular temporal variations observed in the following four geometric angles:

(a) $\psi_1 = 121\lambda_{\text{Pand}} - 118\lambda_{\text{Pro}} - 3\varpi_{\text{Pro}}$,
(b) $\psi_2 = 121\lambda_{\text{Pand}} - 118\lambda_{\text{Pro}} - \varpi_{\text{Pand}} - 2\varpi_{\text{Pro}}$,
(c) $\psi_3 = 121\lambda_{\text{Pand}} - 118\lambda_{\text{Pro}} - 2\varpi_{\text{Pand}} - \varpi_{\text{Pro}}$, and
(d) $\psi_4 = 121\lambda_{\text{Pand}} - 118\lambda_{\text{Pro}} - 3\varpi_{\text{Pand}}$.

where, the subscript $Pro$ refers to Prometheus, respectively, as illustrated in Fig. 4. This behaviour is characteristic of orbits confined within the resonance separatrices. Similarly, Fig. 3(b) shows a detailed view of Pandora's orbit, with several separatrices clearly visible around it.

In the Saturn system, larger moons significantly perturb the orbits of smaller satellites, particularly the Mimas–Tethys pair. Fig. 1 shows the mappings in the neighbourhood of (c) Prometheus and (d) Pandora, which account for these perturbations. Comparing Figs. 1(a) with 1(c), it is evident that higher-order resonant structures have been disrupted, with the domains associated with the Lindblad and corotation resonances being significantly deformed by these perturbations. Similarly, comparing Figs. 1(b) and (d) reveals comparable perturbative effects near Pandora, although the 80:78 resonance remains structurally unchanged.

Analysis of Figs. 3(c) and (d), which incorporate the perturbative effects of the Mimas–Tethys pair, reveals the disappearance of the structures associated with the resonance separatrices observed in Figs. 3(a) and (b). These results suggest that the perturbations from Mimas and Tethys play a significant role in dissipating previously existing higher-order resonances, particularly the 121:118 resonance.

### 3.1. Individual spectra

Studying the individual spectra of the osculating elements of each satellite provides detailed information on their orbital dynamics. Figs. 5(a–c) and 5(d–f) show the spectral analysis of the semi-major axis, eccentricity, and orbital inclination of Prometheus and Pandora, respectively. These spectra exhibit numerous peaks that are not well defined, making the identification of fundamental periods difficult. This behaviour is characteristic of irregular or chaotic orbits.

The eccentricity and inclination spectra of Prometheus show their highest peaks in the secular periods $P_{\Delta\varpi} \approx 2340.57$ days and $P_{\Delta\Omega} \approx 2383.12$ days. For Pandora, these periods differ slightly: $P_{\Delta\varpi} \approx 2319.43$ days and $P_{\Delta\Omega} \approx 2308.79$ days, respectively. The small offsets observed in the secular spectra of Prometheus and Pandora reflect differences in their pericentre and nodal precession rates.

Fig. 6 illustrates the effect of the Mimas–Tethys pair on individual spectra. Several peaks appear on the semi-major axis of Prometheus's spectrum, whereas the eccentricity and inclination spectra show a reduction in the number of peaks. The semi-major axis of Pandora exhibits a reduced number of peaks, whereas the eccentricity and inclination spectra show well-defined peaks, with their main amplitudes demonstrating short-period variations ($P \leq 200$ days).

#### 3.1.1. Individual dynamic power spectra

By fixing Prometheus' eccentricity value at $e_0 = 0.0025$ (red line in Fig. 3(a)) and varying the semi-major axis for Prometheus clones, we can obtain a more precise quantitative knowledge of the frequency distribution in phase-space. This is known as *Individual Dynamic Power Spectra* (IPS). The $y$-axis of the IPS shows the periods corresponding to the peaks in the orbital spectrum, provided that their amplitude exceeds a predefined percentage of the reference amplitude (RA) for each initial condition. Periods longer than 15 days and those associated with the short period and the $J_2$ term have been eliminated, as they are irrelevant to our analysis. The remaining osculating parameters are maintained at constant values (see Table 2).

Fig. 7 shows six IPS for the Prometheus orbital neighbourhood, considering the following VAs: (a,d) semi-major axis; (b,e) eccentricity; and (c,f) orbital inclination. Figs. 7(a–c) consider Pandora to be the only perturbing body, while Figs. 7(d–f) also include the Mimas–Tethys pair. In all cases, an RA = 5% is assumed for the respective parameter.

Fig. 7(a) shows the resonance domain 121:118, which is delimited by blue vertical dashed lines. The red vertical dashed line indicates Prometheus's current semi-major axis (red star in Fig. 3(a)). In this region, the IPS exhibits a dense frequency spectrum, making it difficult to identify the frequencies associated with the resonant angles $\psi_1$, $\psi_2$, $\psi_3$, and $\psi_4$, as well as to determine any fundamental resonant periods. This is the result of the chaotic motion of the system. In Figs. 5(b) and (c), clear horizontal lines correspond to secular periods of eccentricity and inclination: $P_{\Delta\varpi} \approx 2340.57$ days and $P_{\Delta\Omega} \approx 2383.12$ days, respectively (see Section 3.2).

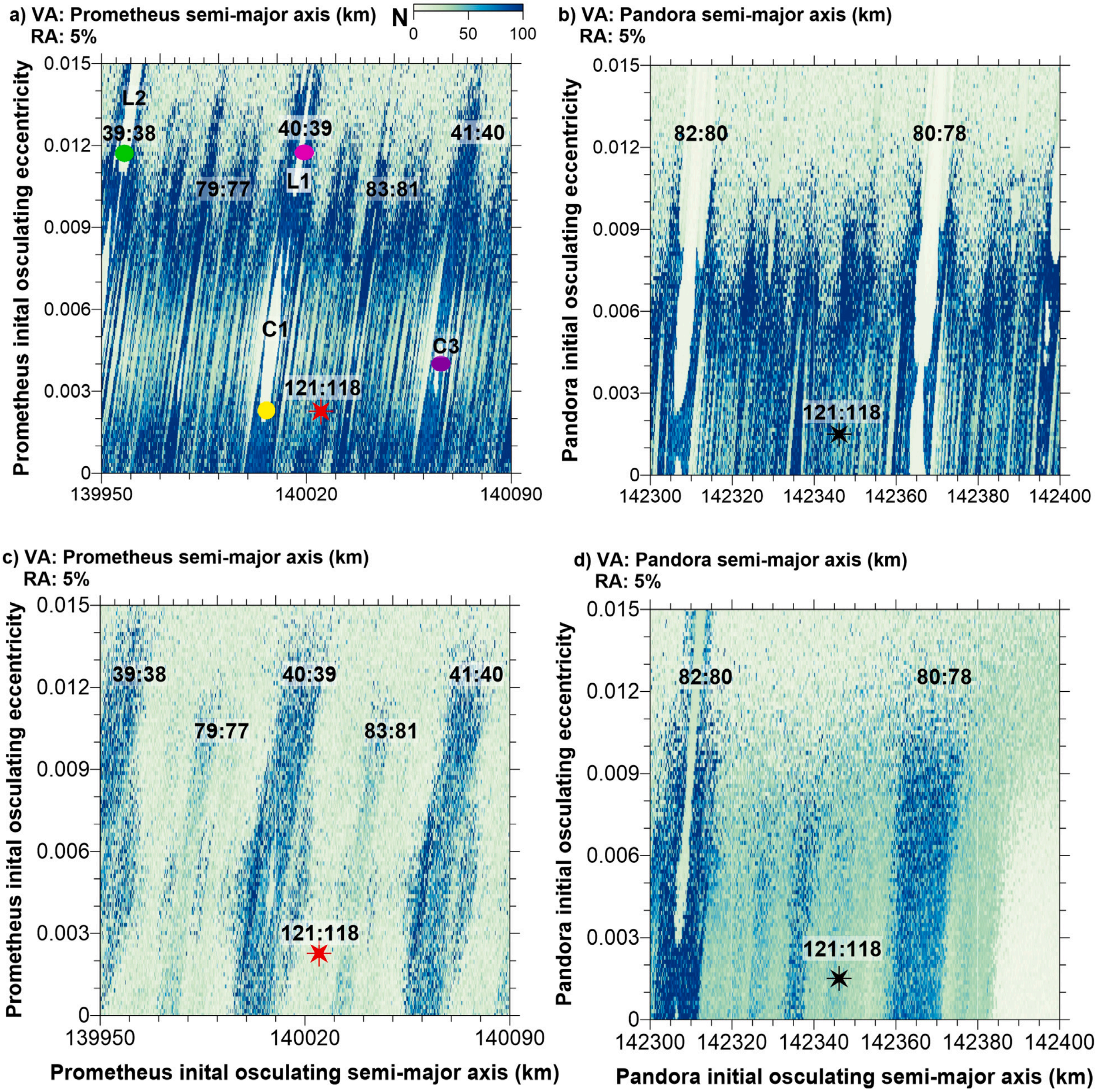

**Fig. 1.** (a) Dynamical map to the region around Prometheus's current orbit (represented by the red star), based on an initial date of 1 January 2000. In this analysis, only Pandora is considered as a disturbing body. **C1**, **C3**, **L1**, and **L2** are the respective corotation and Lindblad zones. The total integration time for the numerical simulation is 188,743.68 days (≈ 516.75 years), with sampling every 0.18 days for each initial condition. **N** is the spectral number. This map shows 71,407 orbits of Prometheus clone satellites, with $\delta a = 0.2$ km, $\delta e = 0.00015$ and RA = 5% using the semi-major axis as the variable of analysis (VA). (b) Dynamical map of Pandora, considering Prometheus as the main perturbing body in this analysis. The black star indicates Pandora's current orbit, with $a_0 \approx$ 142,346.13 km and $e_0 \approx 0.0015$ (1 January 2000). Two vertical structures are observed, associated with the 80:78 and 82:80 Pandora–Prometheus resonances. (c) and (d) show the same mappings as (a) and (b), respectively; however, the effects of the disturbance caused by the Mimas–Tethys satellite pair have been included. The presence of this pair is found to induce significant perturbations in the structures associated with the Lindbald and corotation resonances previously identified in (a). In contrast, the 80:78 resonant structure remains stable in (d), whereas the 82:80 resonance exhibits changes in its configuration.

Including the Mimas–Tethys perturbation eliminates the resonant domain (Fig. 7(a)). However, the secular periods identified in Figs. 7(b) and (c) are still evident in Figs. 7(e) and (f). This suggests that the secular perturbations between Prometheus and Pandora persist even when the Mimas–Tethys pair is included. This indicates that the influence of these additional perturbations on the secular dynamics of Prometheus's orbit is minimal.

Fig. 8 is analogous to Fig. 7, but now shows the IPS for the orbital neighbourhood of Pandora. In this case, the perturbations are due to Prometheus alone in 8(a-c), and to Prometheus in combination with the Mimas–Tethys pair in 8(d–f). As in Fig. 7(d), the resonant domain in Fig. 8(d) is absent, giving way to several vertical peaks without a clear physical interpretation. However, Figs. 8(b) and (c) reveal the corresponding secular periods, $P_{\Delta\varpi} \approx 2319.43$ days, and $P_{\Delta\Omega} \approx 2308.79$ days. These secular periods are also evident in Figs. 6(e) and (f).

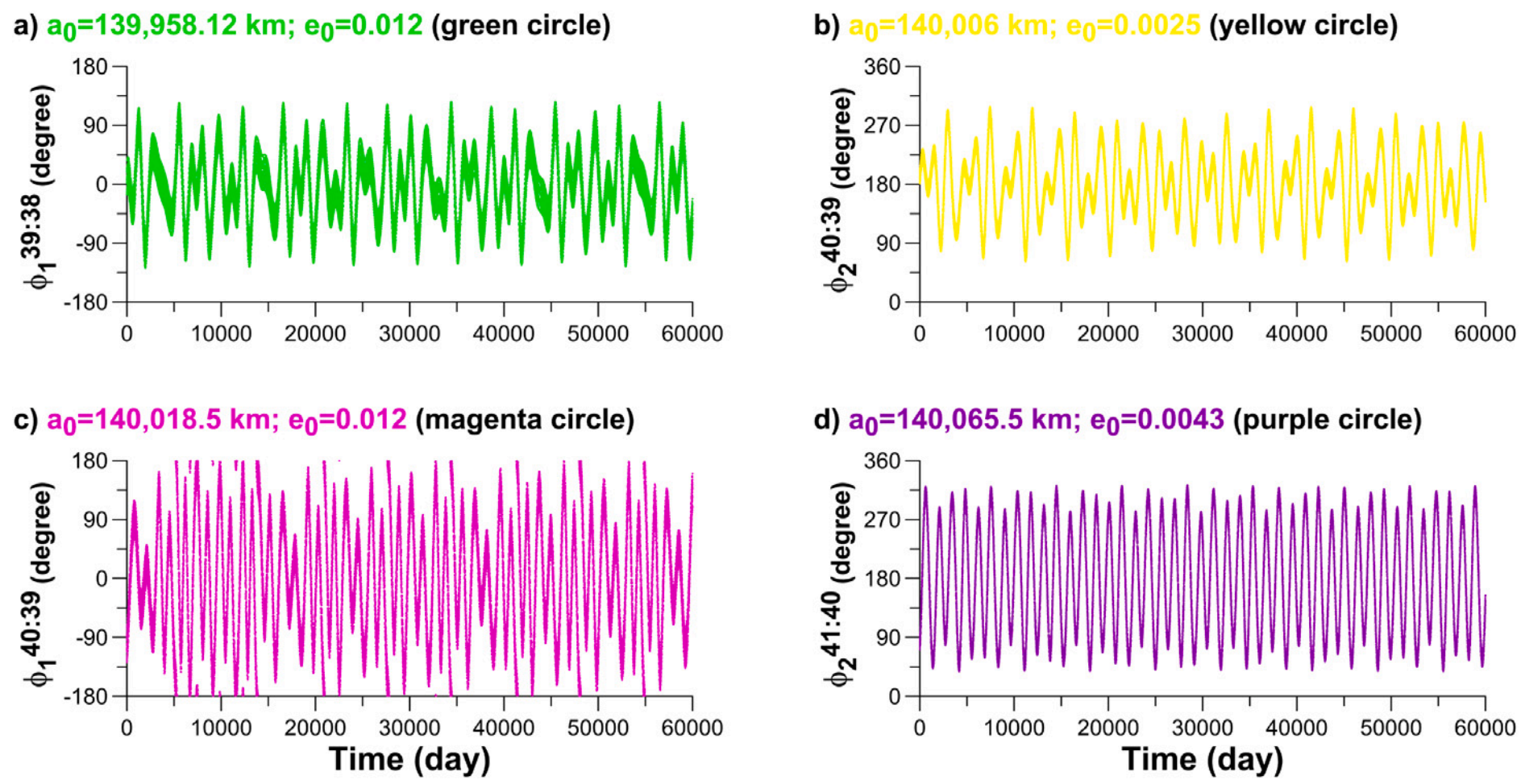

**Fig. 2.** Temporal variation of the angles $\phi_1^{m+1:m}$, and $\phi_2^{m+1:m}$, where $m = 38$, 39, and 40, is associated with Lindblad and corotation resonances, respectively. For the Prometheus clone orbits with the initial conditions shown in Fig. 1(a) (coloured circles), we observe the following: (a,c) oscillation around 0 for the Lindblad angle and (b,d) oscillation around 180° for the corotation angle.

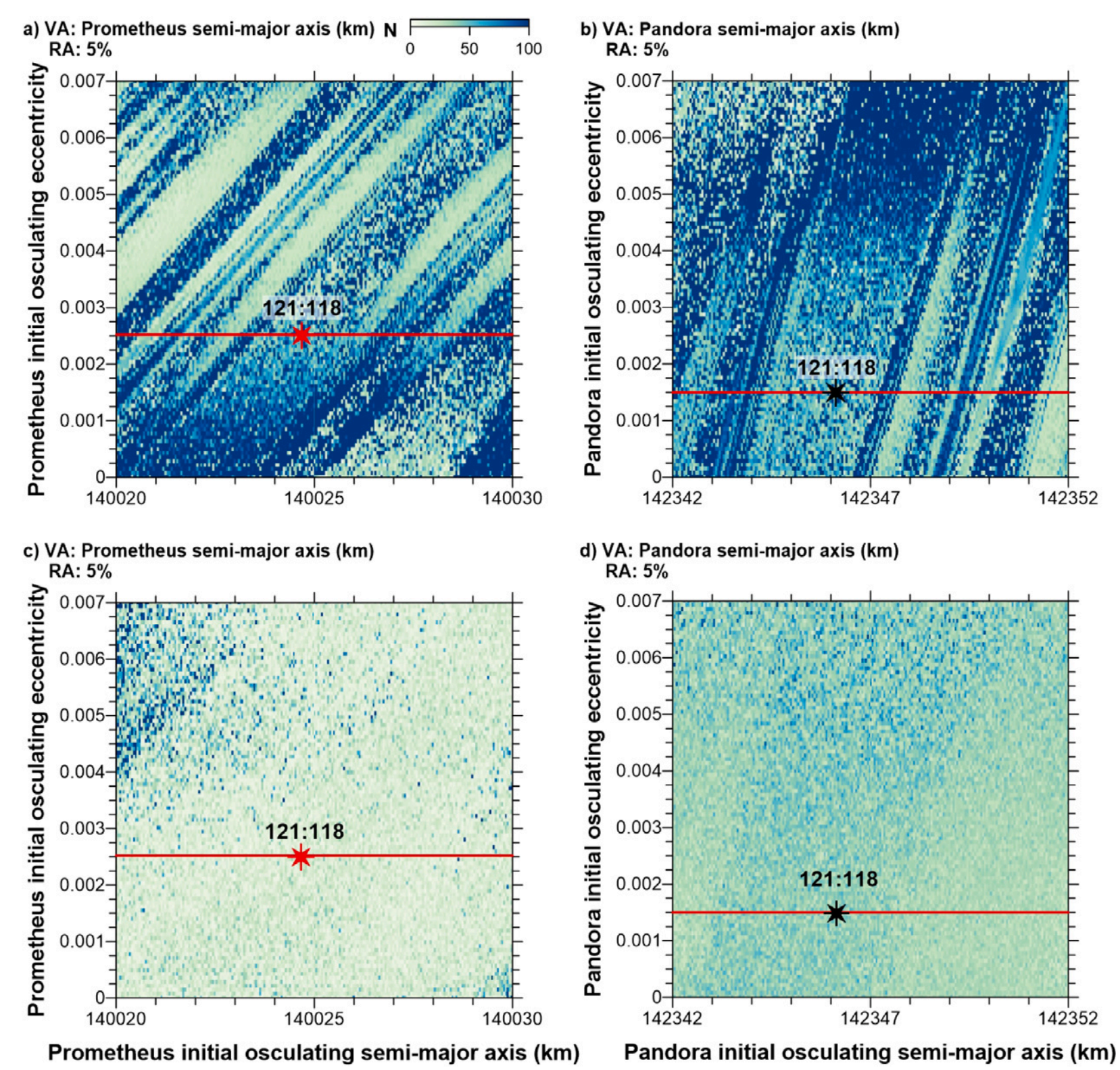

**Fig. 3.** Dynamical map provides a detailed view of the orbital neighbourhood of Prometheus and Pandora, as shown in Fig. 1.(a) Region near the current orbit of Prometheus (red star). The 121:118 resonance domain appears as a narrow band (high value of **N**), indicating the irregular nature of the orbits within its domain. The current orbit of Prometheus lies on the separatrices of the 121:118 resonance with Pandora. (b) Similar to Prometheus, Pandora (black star) is surrounded by several separatrices associated with the 121:118 resonance, which is characteristic of regions exhibiting irregular dynamical behaviour. (c) Same as (a), but including the perturbative effects of the Mimas–Tethys pair. (d) Similar to (b), but with the Mimas–Tethys perturbations taken into account. Including these perturbations contributes to the destruction of the separatrices observed in (a) and (b).

In contrast, the secular periods identified in Figs. 7(e), (f), 8(e) and (f) correspond to ≈ 6.4 years. This is very close to the period associated with the anti-alignment of Prometheus and Pandora (≈ 6.2 years). This suggests that the secular perturbation between the two satellites is linked to the anti-alignment period of their pericentres.This topic is covered in more detail in Section 3.2.1.

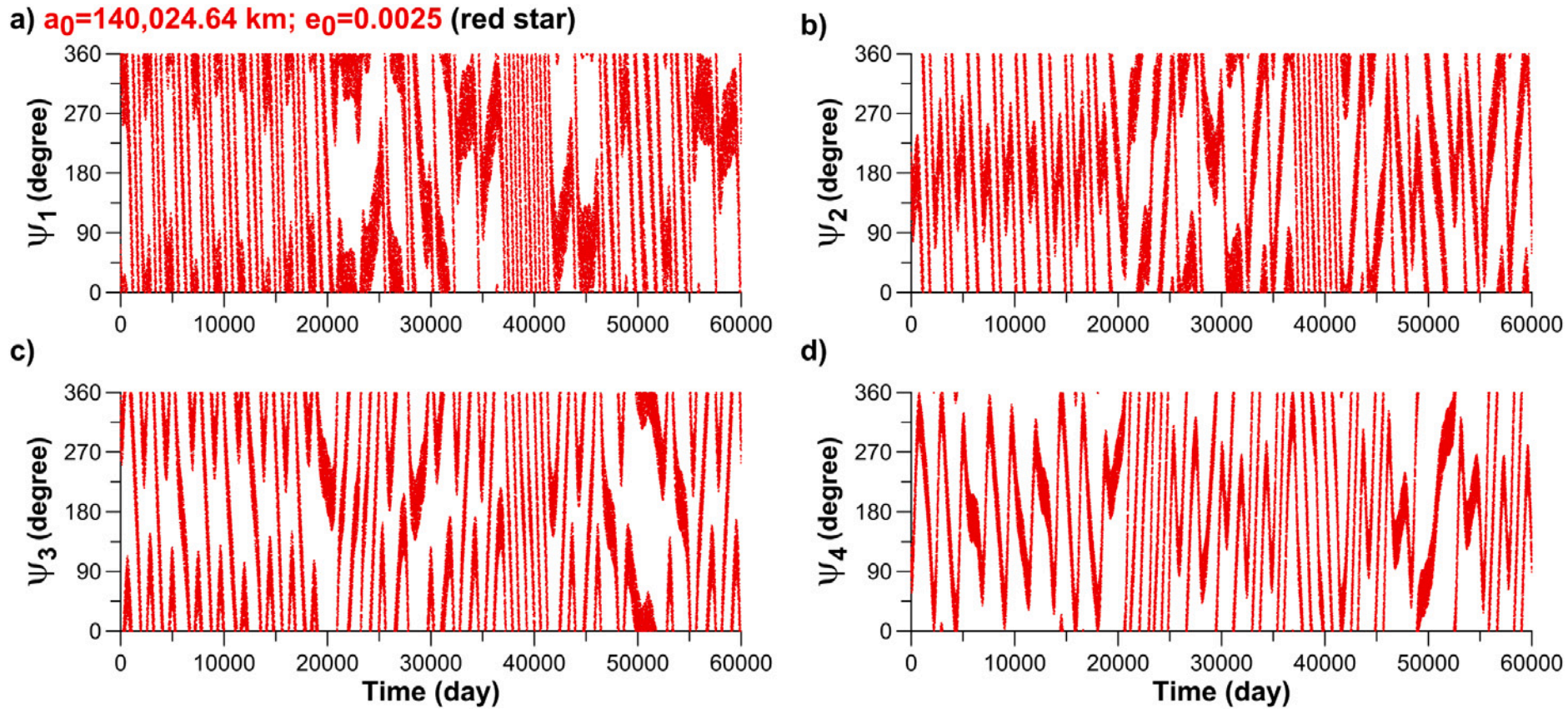

**Fig. 4.** Temporal variation of the geometric angles (a) $\psi_1$, (b) $\psi_2$, (c) $\psi_3$ and (d) $\psi_4$ related to the 121:118 resonance with Pandora, obtained by numerically integrating the current position of Prometheus (red star in Fig. 1) and Pandora (black star in Fig. 1) using the MERCURY package. We observed the alternation between oscillations and circulations, which is typical behaviour for orbits internal to the separatrices of a resonance.

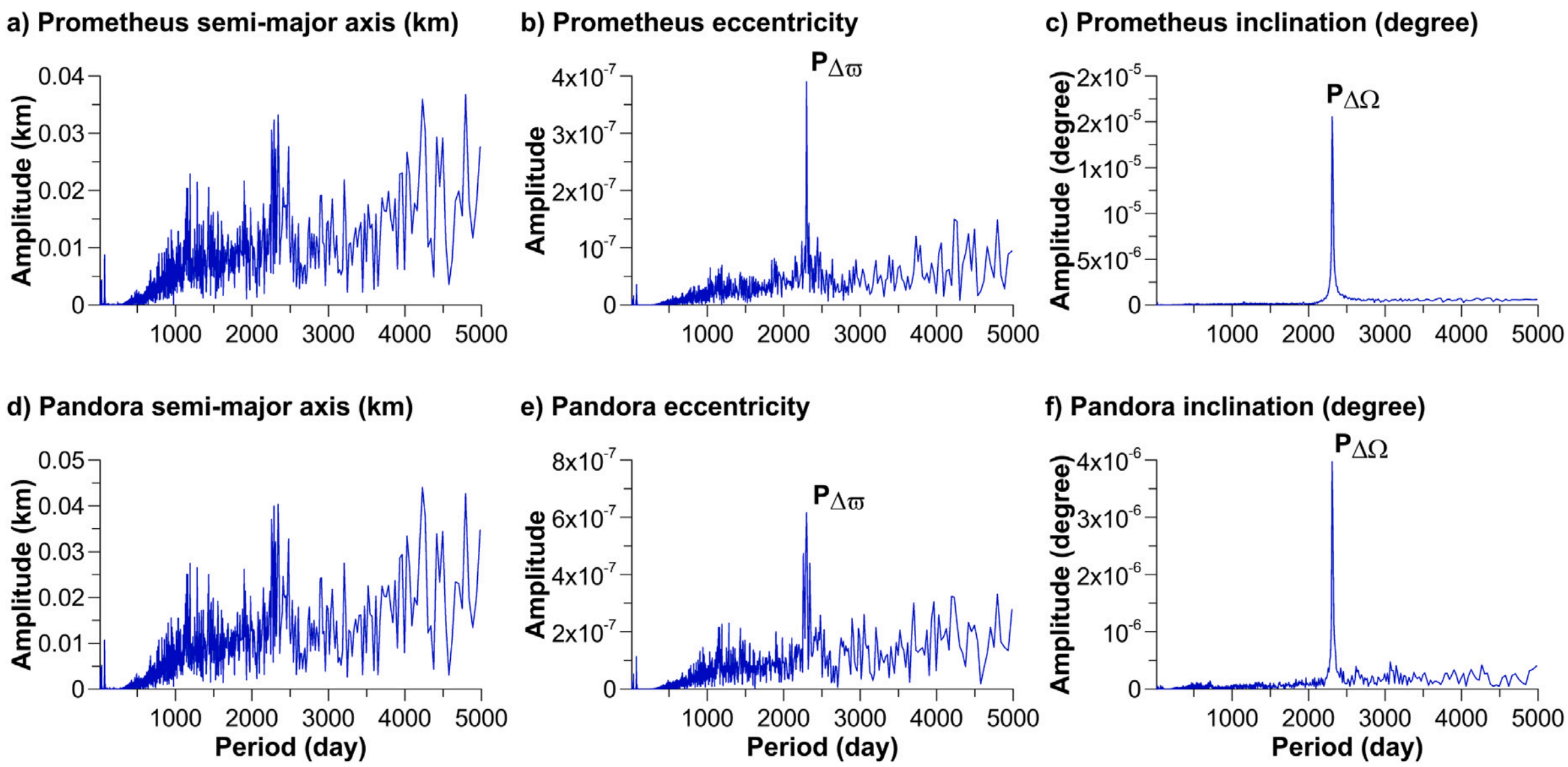

**Fig. 5.** Individual spectra for Prometheus's and Pandora's current osculating orbits on 1 January 2000. (a,b,c) correspond to Prometheus's semi-major axis, eccentricity, and orbital inclination, respectively, while (d,e,f) show the same parameters for Pandora. The high number of peaks observed in the spectra is characteristic of irregular orbits and is related to the satellites' current orbital positions (obtained similarly to Fig. 3). For Prometheus, the secular periods for eccentricity ($P_{\Delta\varpi} \approx 2340.57$ days) and inclination ($P_{\Delta\Omega} \approx 2383.12$ days) are visible in (b) and (c), respectively. For Pandora, the corresponding secular periods are $P_{\Delta\varpi} \approx 2319.43$ days and $P_{\Delta\Omega} \approx 2308.79$ days, visible in (e) and (f). All spectra were obtained by numerical integration over 1720 years with a 0.06 day interval.

### *3.2. Secular and $J_2$ short-term variations on the current orbits of Prometheus and Pandora*

To accurately characterise the dynamical behaviour of the Prometheus–Pandora system, we adopt a system consisting of Saturn, Prometheus and Pandora. In this approach, deviations from the classical Newtonian formulation arise exclusively from Saturn's oblateness, which is accounted for by including its dominant zonal harmonic, $J_2$. The osculating orbital elements exhibit rapid, pronounced oscillations associated with the zonal harmonics of Saturn ($J_2$, $J_4$, and $J_6$), which can mask the underlying secular dynamics. To suppress these short-period effects, we employed the geometric element algorithm proposed by Renner and Sicardy (2006), which provides a filtered representation of the orbital elements that is practically free from high-frequency perturbations. This procedure is illustrated in Figs. 9 and 10, which compare the osculating (black curves) and geometric (green curves) elements: (a, b) the semi-major axis; (c, d) the eccentricity; and (e, f) the orbital inclination for Prometheus and Pandora, respectively. For Prometheus, the dominant $J_2$ oblateness term causes oscillations in the osculating eccentricity, with an amplitude of around 0.0045 (see Fig. 9(c)), whereas the geometric eccentricity remains smooth, reflecting the orbit's secular evolution. Similar behaviour is observed for Pandora, where the osculating eccentricity exhibits larger amplitude variations of about 0.007 (Fig. 10(c)). In both cases, the geometric elements effectively remove the short-period noise introduced by Saturn's gravitational harmonics.

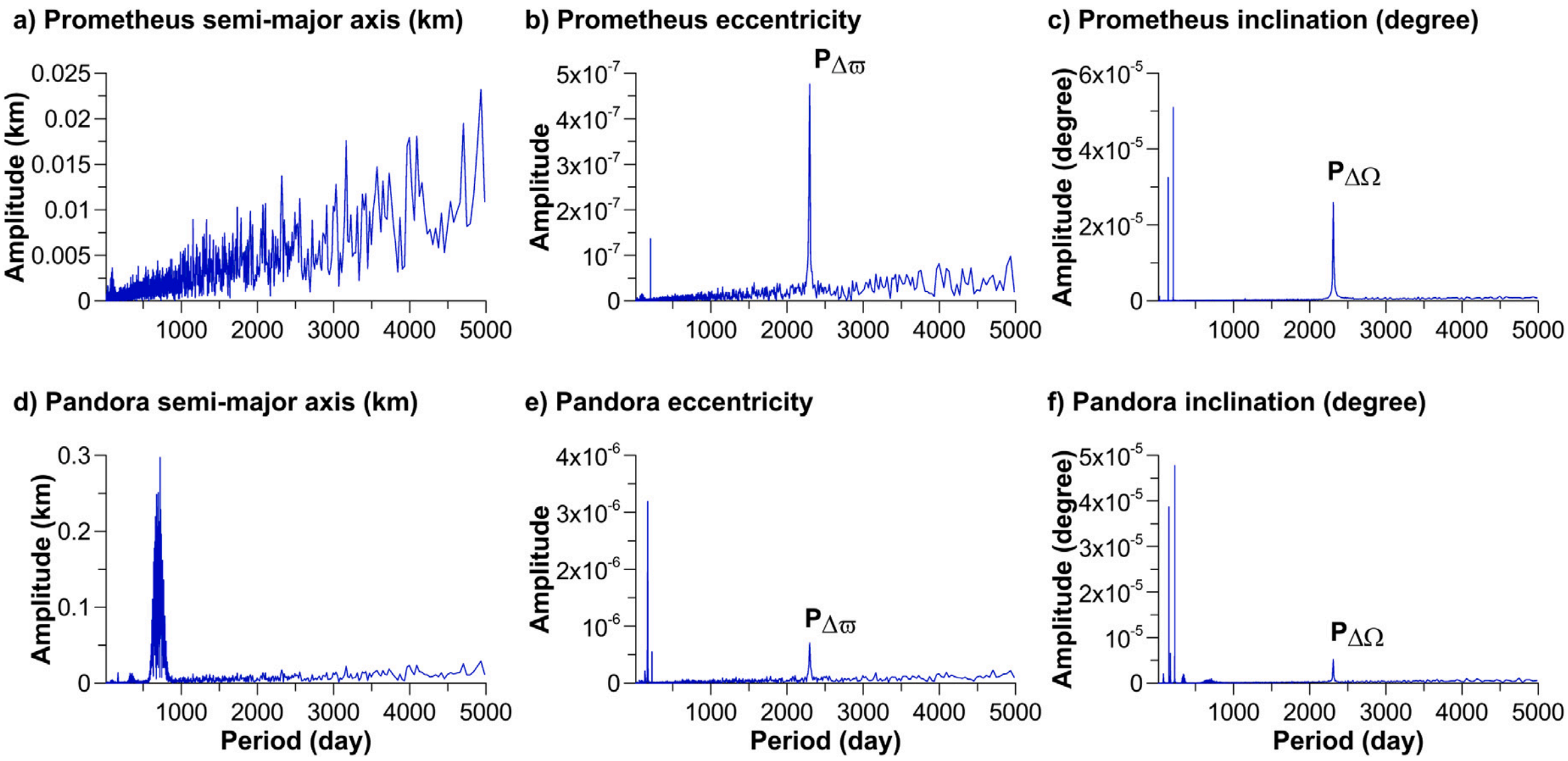

**Fig. 6.** Spectral analysis reveals the perturbative effects of the Mimas–Tethys pair on the orbits of Prometheus and Pandora, as shown in Fig. 5. (a) shows the semi-major axis of Prometheus, which exhibits multiple spectral peaks. In contrast, the spectra of its (b) eccentricity and (c) inclination reveal a reduction in the number and amplitude of peaks compared to previous findings. The semi-major axis spectrum of Pandora (d) shows fewer peaks than Fig. 5(d), while the eccentricity (e) and inclination (f) spectra reveal well-defined periods associated with short-period variations ($P \leq 200$ days).

Variations in eccentricity and inclination can be described more precisely in terms of two secular angular variables that capture their long-term dynamical evolution: the difference in the longitudes of the pericentre, $\Delta\varpi_{\text{Pand-Pro}} \equiv \varpi_{\text{Pand}} - \varpi_{\text{Pro}}$, and the difference in the longitudes of the ascending node, $\Delta\Omega_{\text{Pand-Pro}} \equiv \Omega_{\text{Pand}} - \Omega_{\text{Pro}}$ (Brouwer and Clemence, 1966). These angular variables are shown in Fig. 9(g–j). According to Callegari et al. (2021), the fundamental frequencies associated with these angles correspond to the mutual secular modes of the system, arising from the combined effects of mutual gravitational perturbations and the zonal harmonics of Saturn.

For the Prometheus–Pandora pair, the corresponding secular periods are found to be $P_{\Delta\varpi} \approx 2340.57$ days and $P_{\Delta\Omega} \approx 2383.12$ days. These long-term periods are clearly identifiable in Fig. 9, which shows how eccentricity, inclination, and secular angles $\Delta\varpi_{\text{Pand-Pro}}$ and $\Delta\Omega_{\text{Pand-Pro}}$, together with their respective secular components.

Fig. 9 highlights the distinct roles that secular and short-period perturbations play in Prometheus' orbital evolution. Although the $J_2$ term produces noticeable oscillations in the osculating eccentricity (Fig. 9(c)), its effect on the inclination is very small (Fig. 9(e)). In contrast, the geometric inclination is essentially unaffected by short-period perturbations, with its dominant variations arising from the secular component, as evidenced in Figs. 9(b) and (e). A similar comparison for Pandora is shown in Fig. 10, where the influence of the $J_2$ harmonic again manifests itself primarily as large-amplitude oscillations in the osculating eccentricity and as weaker modulations in the inclination.

In order to determine the secular effects on eccentricity and inclination quantitatively, we examine the projection of Prometheus's osculating, geometric, and secular orbits onto the planes ($e_{\text{Pro}}\cos(\Delta\varpi_{\text{Pand-Pro}})$, $e_{\text{Pro}}\sin(\Delta\varpi_{\text{Pand-Pro}})$), and ($I_{\text{Pro}}\cos(\Delta\Omega_{\text{Pand-Pro}})$, $I_{\text{Pro}}\sin(\Delta\Omega_{\text{Pand-Pro}})$).

The red curves in Fig. 11(a,c) represent the projection of the analytical secular theory solution for eccentricity and inclination (see Murray and Dermott (1999), Chapter 7), while the black and green curves represent the osculating and geometric eccentricity and inclination of Prometheus in the respective polar planes defined above. Figs. 11(a) and (c) show that the forced term for eccentricity and inclination is negligible and that the results are consistent with the variable spectra shown in Figs. 5(b) and (c), which exhibit small amplitudes.

Similarly, the projections of the secular, osculating, and geometric orbits shown in Figs. 11(a) and (d) reveal the absence of a forced component in Pandora's eccentricity and inclination, respectively.

### 3.2.1. Dynamical interpretation of the 6.2-year secular period: the Prometheus–Pandora anti-alignment

In addition to the numerical results presented, it is possible to analytically estimate the period associated with the secular circulation of the difference in the longitudes of the pericentre, $\Delta\varpi_{\text{Pand-Pro}}$, based on the classical perturbation theory. In systems with oblate central bodies such as Saturn, the pericentre precession is strongly influenced by the non-spherical gravitational potential, particularly by the dominant harmonic $J_2$. The mean precession rate induced by this term can be expressed as follows:

$$\dot{\varpi} = \frac{3}{2}\frac{J_2 n}{(1-e^2)^2}\left(\frac{R_S}{a}\right)^2\left(2 - \frac{5}{2}\sin(i)\right)^2, \tag{6}$$

$$\dot{\Omega} = -\frac{3}{2}\frac{J_2 n}{(1-e^2)^2}\cos(i), \tag{7}$$

where $n$, $a$, $e$, and $i$ represent the mean-motion, satellite's semi-major axis, eccentricity, and orbital inclination, respectively, and $R_S$ is Saturn's equatorial radius. This expression results from expanding the disturbing function for the $J_2$ perturbations (see Danby, 1988). When the orbit is close to the celestial equator of the main body and the eccentricity is close to zero, Eqs. (6) and (7) can be approximated by:

$$\dot{\varpi} \approx \frac{3}{2}J_2 n\left(\frac{R_S}{a}\right)^2, \qquad \dot{\Omega} \approx -\frac{3}{2}J_2 n\left(\frac{R_S}{a}\right)^2.$$

Applying this approximation to the mean values of Prometheus $a_{\text{Pro}} \approx 139{,}380$ Km and Pandora $a_{\text{Pand}} \approx 141{,}720$ km yields the precession rates: $\dot{\varpi}_{\text{Pro}} \approx 134.42$ days, $\dot{\varpi}_{\text{Pand}} \approx 142.49$ days. The difference in these rates, $\Delta\dot{\varpi}_{\text{Pro-Pand}} = |\dot{\varpi}_{\text{Pand}} - \dot{\varpi}_{\text{Pro}}|$, determines the rate at which $\Delta\varpi$ circulates, and the corresponding secular period is given by:

$$P_{\text{secular}} = \frac{1}{\Delta\dot{\varpi}_{\text{Pro-Pand}}} \approx 6.2\,\text{years}.$$

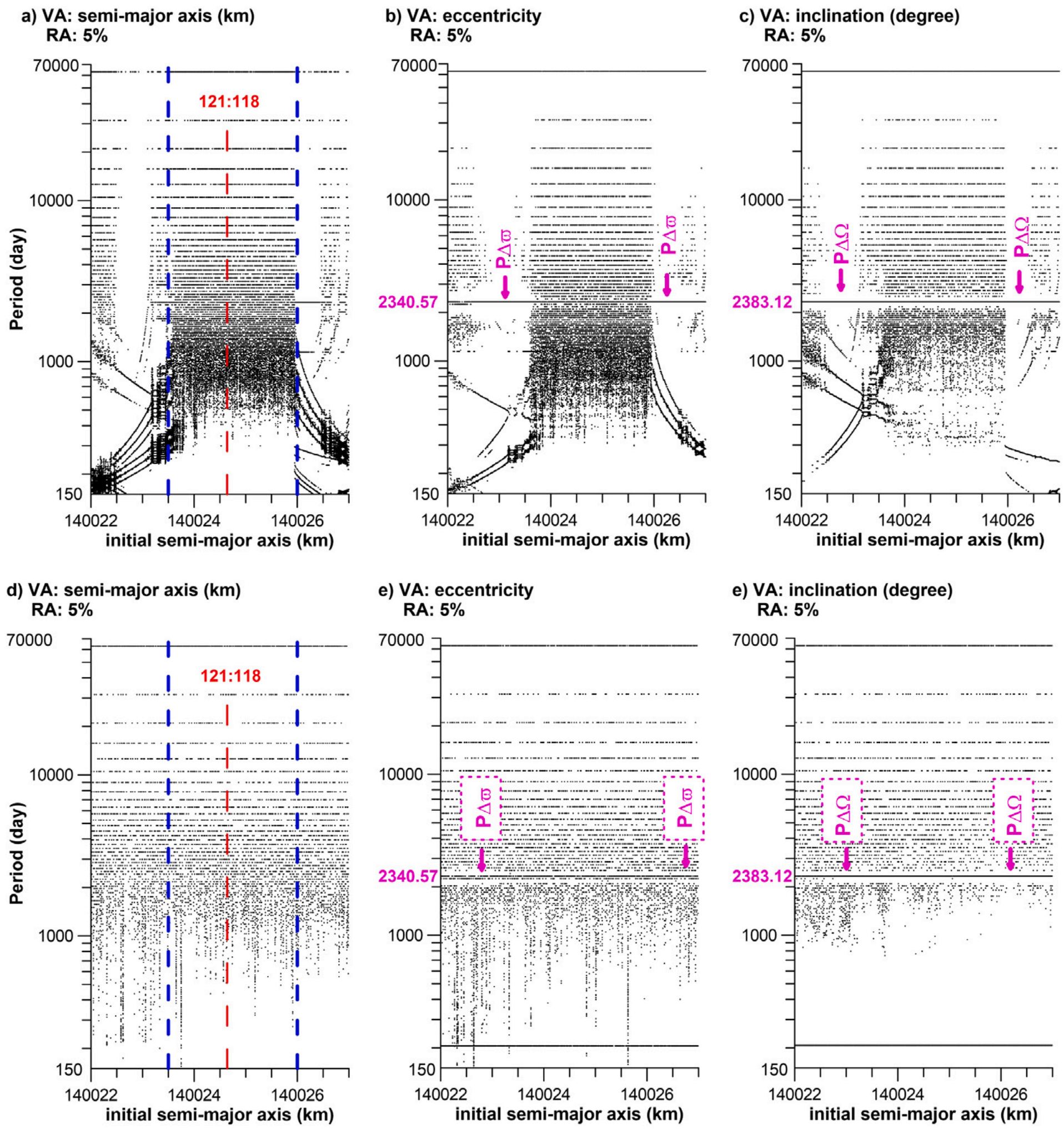

**Fig. 7.** IPS for 2000 Prometheus clones considering variations in: (a,d) semi-major axis, (b,e) orbital eccentricity, and (c,f) orbital inclination. In (a), the blue vertical dashed lines indicate the 121:118 resonance domain with Pandora, and the red vertical dashed line shows the current value of Prometheus's semi-major axis. This region exhibits a dense distribution of frequencies, which makes it difficult to identify the fundamental periods. (b) and (c) show the secular periods associated with eccentricity ($P_{\Delta\varpi} \approx 2340.57$ days) and inclination ($P_{\Delta\Omega} \approx 2383.12$ days), respectively. (d), (e), and (f) incorporate the perturbative effects of the Mimas–Tethys pair, as modelled in Figs. 3(c) and (d). The inclusion of these perturbations has the following effects: (d) the destruction of the 121:118 resonant domain between Pandora and Prometheus, and (e,f) the preservation of the secular periods identified in (b) and (c). Our results suggest that the presence of the Mimas–Tethys pair does not affect the secular disturbances between Prometheus and Pandora. The integration time for each orbit is 172.25 years, with a time sampling at 0.06 days intervals; only periods greater than 150 days are shown.

This theoretical value is in agreement with the numerical results shown in Fig. 9(h), which is based on the temporal variation of $\Delta\varpi_{Pand-Pro}$ obtained from numerical integrations. In this figure, a continuous and regular circulation of the difference between the pericentres is observed, with a period compatible with the analytically estimated 6.2 years. The geometric curve, in particular, highlights this behaviour more clearly, as it removes the short-term component associated with the rapid osculating precession of Prometheus's pericentre.

Furthermore, Fig. 12 shows the spectral analysis for $\Delta\varpi$, where the dominant peak corresponds to a period of approximately 6.2 years (2264.55 days), This confirms that the secular variation identified analytically corresponds to the anti-alignment of Prometheus–Pandora. This periodic anti-alignment acts as a natural protective mechanism between the orbits of Prometheus and Pandora, preventing close encounters.

Similar results were reported by Goldreich and Rappaport (2003b) and Cooper and Murray (2004), who identified the same period related to the periodic anti-alignment of the orbits, with implications for close encounters and chaotic dynamics in the system. This consistency between theory and simulation strengthens the validity of the secular approach in describing the orbital architecture of Prometheus and Pandora.

## 4. Quantifying the effective stability of Prometheus and Pandora

### 4.1. Limitations of classical chaos indicators

Chaos detection tools, such as Lyapunov exponents and spectral numbers, are widely used to distinguish between regular and chaotic orbits in the phase space of dynamical systems. However, in more complex systems, these methods show limitations in identifying phenomena such as resonant captures and slow or confined chaos (Tsiganis et al., 2002; Winter et al., 2010), as well as in establishing a clear relationship between chaoticity and long-term macroscopic stability (Morbidelli and Froeschlé, 1995). Moreover, variational and spectral methods can produce artefacts and fail to accurately capture the fine structure of the phase space (Barrio et al., 2009).

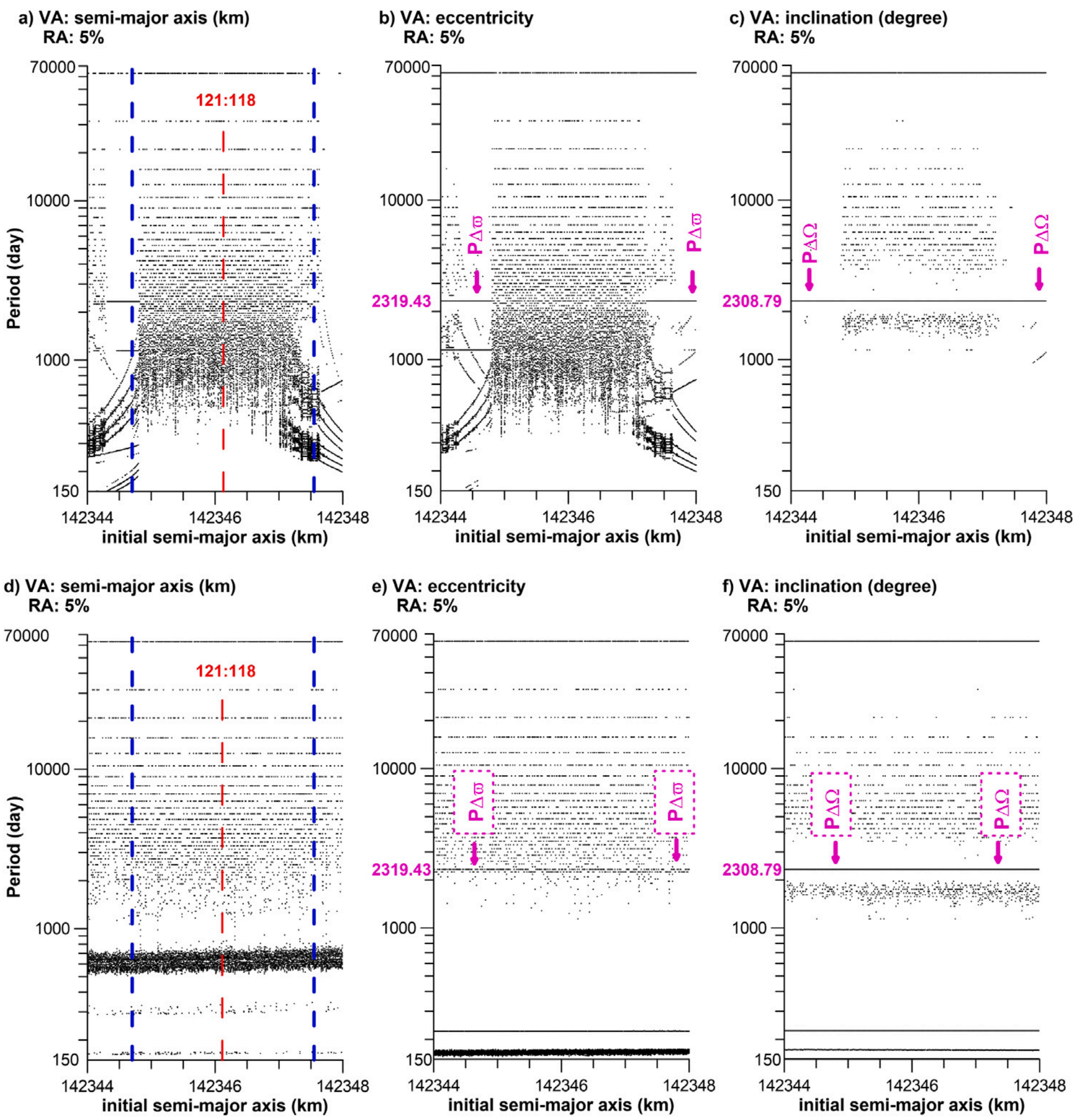

**Fig. 8.** IPS for Pandora's orbital neighbourhood was constructed by fixing its initial eccentricity at $e_0 = 0.0015$ (red line in Figs. 3(b) and (d)). (a), (b) and (c) show the IPS for variations in the semi-major axis, eccentricity, and inclination, respectively, when Pandora is only perturbed by Prometheus. (a) clearly illustrates the domain associated with the 121:118 resonance. (b) and (c) reveal the corresponding secular periods, $P_{\Delta\varpi} \approx 2319.43$ days in eccentricity and $P_{\Delta\Omega} \approx 2308.79$ days for inclination. These are slightly different to the periods obtained for Prometheus. (d) shows that the resonant domain is no longer present, giving way to vertical peaks without clear physical meaning. These secular periods persist in (e) and (f), indicating that the secular dynamics between Pandora and Prometheus remain dominant despite the inclusion of Mimas–Tethys perturbations.

In this context, Guimarães and Michtchenko (2023) explored the methodology to investigate the relationship between the diffusion of dynamical frequencies and orbital stability. This approach was subsequently applied to the Saturnian system, with a particular focus on Atlas, by Ceccatto et al. (2025). Their findings demonstrate that low levels of frequency drift, known as 'confined chaos', can coexist with long-term orbital confinement, even in regions that would typically be categorised as chaotic based on traditional indicators.

### 4.2. Frequency-domain analysis and wavelet methods

The study of chaotic diffusion often involves analysing how action variables, such as the semi-major axis and eccentricity, evolve over time in the action domain (Froeschlé et al., 2005; Cordeiro, 2005; Cachucho et al., 2010). In contrast, analysing diffusion in the frequency domain has been shown to more precisely capture the fine structure of the phase space (Laskar, 1990). Laskar (1993) identified a relationship between the spatial and temporal diffusion of independent frequencies, expressed by a diffusion equation:

$$\nabla^2 f(\vec{x}, t) \propto \frac{\partial}{\partial t} f(\vec{x}, t). \tag{8}$$

As illustrated by individual spectra and IPS in Figs. 5–8, periodic orbits exhibit constant independent frequencies ($\partial f/\partial t = 0$), while chaotic orbits show frequency drift over time ($|\partial f/\partial t| > 0$), resulting in sharp or broadened spectral lines, respectively. Measurement of this drift offers valuable insight into the intricate dynamical behaviour of the system.

Several spectral methods have been developed to track the time evolution of independent frequencies, including the Numerical Analysis of Fundamental Frequencies (NAFF) (Laskar, 1993) and the Frequency-Modified Fourier Transform (FMFT) (Šidlichovskỳ and Nesvornỳ, 1996). In this study, however, we use the Continuous Wavelet Transform (CWT), introduced by (Michtchenko and Nesvorny, 1996) as the Wavelet Analysis Method (WAM). This technique is particularly effective in distinguishing between weak and strong chaotic behaviour, accurately resolving independent frequencies, and isolating long-period components—thus overcoming the limitations of previous approaches.

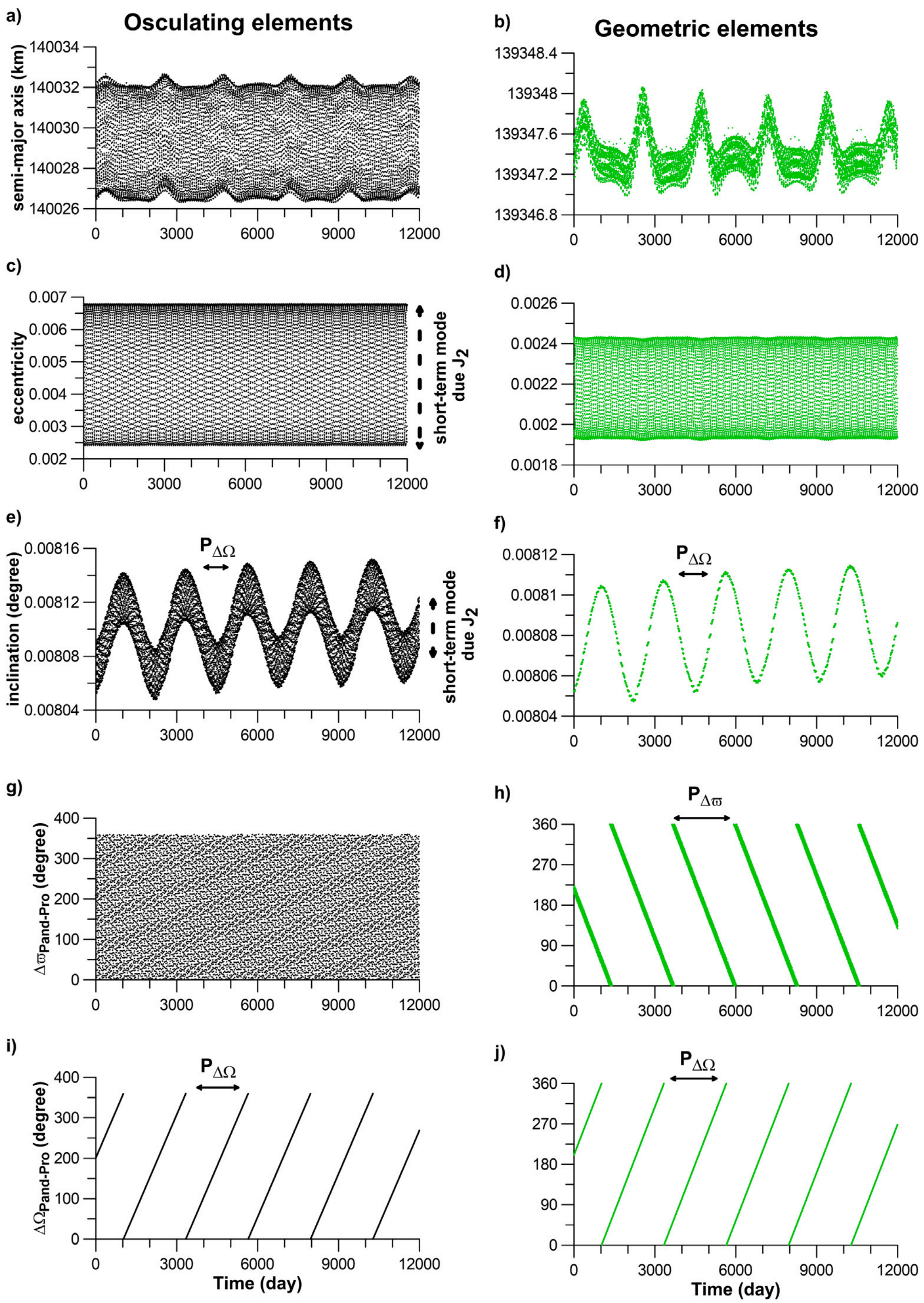

**Fig. 9.** Comparison between the osculating (on the left) and geometric (on the right) elements for the temporal variation of (a,b) semi-major axis, (c,d) eccentricity and (e,f) orbital inclination (degree) relative to Saturn's equator. In (c), the amplitude of $\approx 0.0045$ is effect of the short-term component caused by $J_2$ on the osculating eccentricity. The rapid circulation of the Prometheus's pericentre affects the (g) osculating $\Delta\varpi_{Pand-Pro}$, making its interpretation difficult. (h) geometric $\Delta\varpi_{Pand-Pro}$, (i) osculating and (j) geometric $\Delta\Omega_{Pand-Pro}$. This simulation used the starting date 1 January 2000 sampled every 1 day. The satellites included Prometheus and Pandora, along with the terms $J_2$, $J_4$, and $J_6$.

### 4.3. Diffusion coefficients and stability classification

To quantify the strength of dynamical diffusion, we define a diffusion coefficient based on the relative change in each independent frequency over time, following (Robutel and Gabern, 2006):

$$\delta f_i = \frac{|\ \bar{f}_i^1 - \bar{f}_i^2\ |}{\bar{f}_i^1},$$

where $\bar{f}_i^1$ and $\bar{f}_i^2$ are the average values of the frequency $f_i$ in the first and second halves of the integration time, respectively. This formulation is particularly suitable for use with WAM-derived frequency data and effectively handles the different scales among the frequencies involved (e.g. $n \sim 0.6$ days vs. $f_\varpi$ and $f_\Omega > 100$ days).

The calculated diffusion coefficients are analysed using the classification criteria proposed by Marzari et al. (2003) and Robutel and

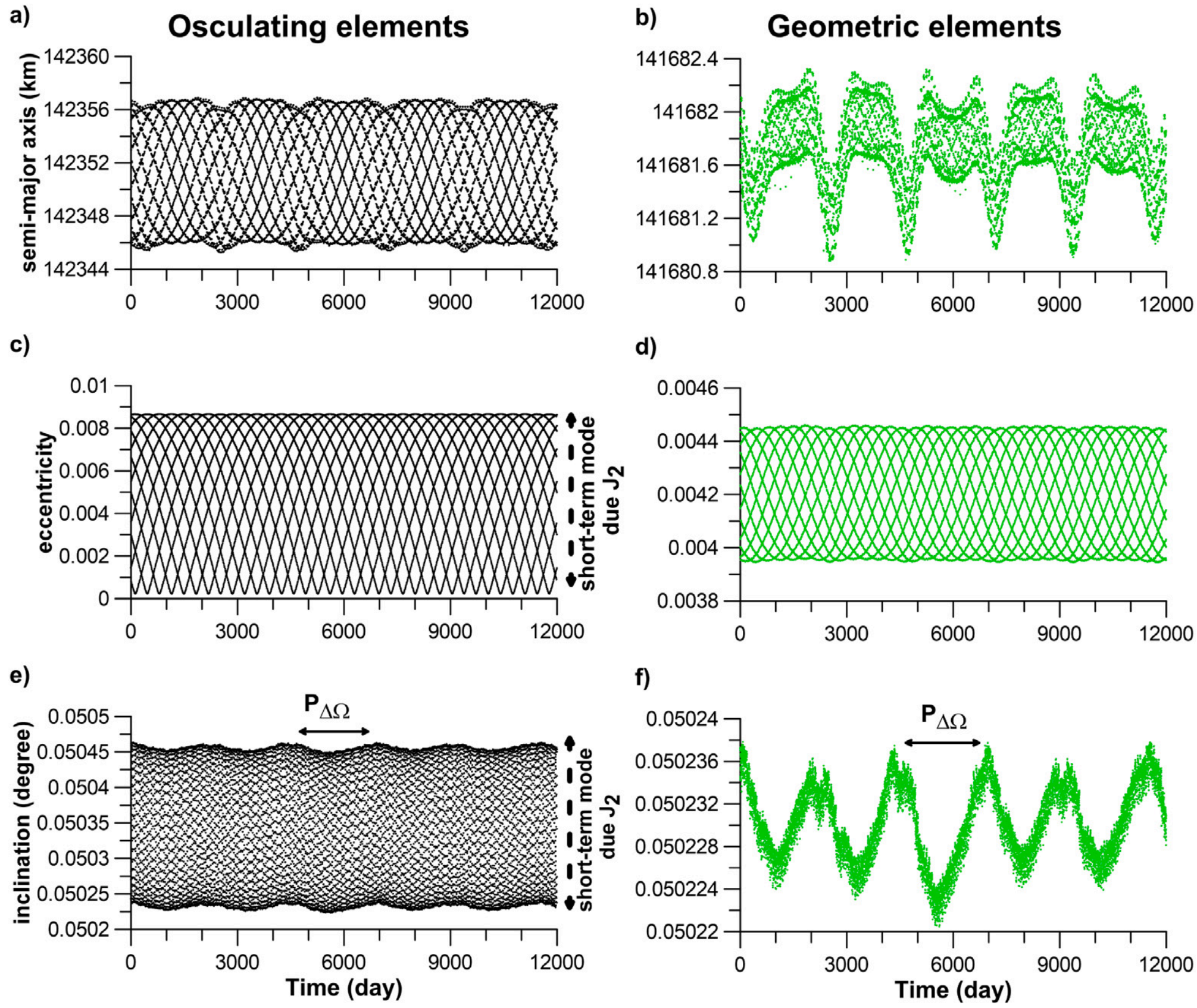

**Fig. 10.** A comparison of the osculating (black) and geometric elements for the semi-major axis (a,d), eccentricity (b,d), and orbital inclination (c,f) of Pandora. As can be seen, the $J_2$ component produces a variation in Pandora's eccentricity of approximately 0.007, as well as a small variation in inclination.

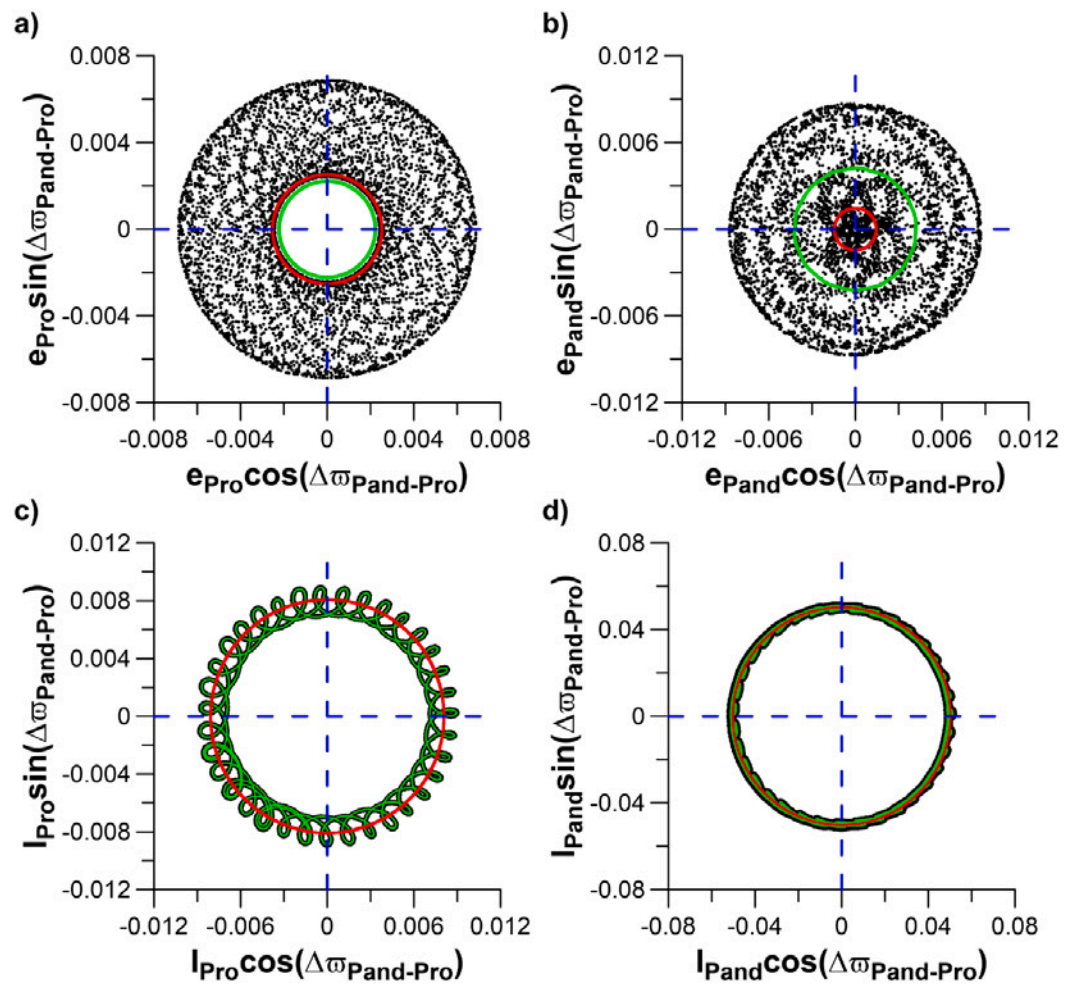

**Fig. 11.** The osculating (black curve), geometric (green curve) and secular (red curve) eccentricities and inclinations are shown for Prometheus (a, c) and Pandora (b, d), respectively. The red curve was obtained using classical secular theory, incorporating Saturn, Prometheus and Pandora, as well as the coefficients $J_2$ and $J_4$. These figures show the polar projection and the absence of the forced term for these variables. The effects of the $J_2$ term on these variables can be seen in Figs. 9(c,e) and 10(c,e), respectively.

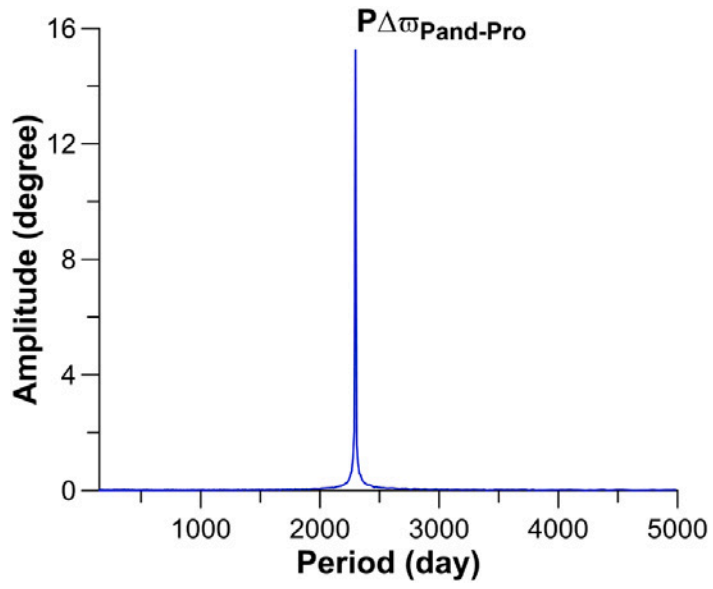

**Fig. 12.** Spectrum of the variable $\Delta\varpi$, showing a dominant secular period of approximately $P_{\Delta\varpi} \approx 6.2$ years (2264.55 days), associated with the anti-alignment between the orbits of Prometheus and Pandora.

Gabern (2006). Although these thresholds are empirical, they are widely used in the literature on frequency diffusion and weakly chaotic Hamiltonian systems (e.g. Marzari et al., 2003; Robutel and Gabern, 2006). In particular, values of $\delta f_i \leq 10^{-5}$ are typically associated with nearly invariant quasiperiodic tori, for which the fundamental frequencies remain effectively constant over the integration interval, signifying regular motion. In contrast, diffusion levels approaching $\delta f_i \geq 10^{-1}$ indicate rapid frequency drift and strong stochastic transport across resonant layers, which are characteristic of strongly chaotic trajectories. Intermediate values ($10^{-5} < \delta f_i < 10^{-1}$) correspond to weakly chaotic motion or quasi-periodicity, in which the orbit undergoes slow modulation of the fundamental frequencies while remaining dynamically confined over long time intervals.

### 4.4. Application to the Prometheus–Pandora system

In this study, we analyse the Prometheus–Pandora system in two configurations: (i) as an isolated pair and (ii) including perturbations from Mimas and Tethys. These configurations are chosen to evaluate how external gravitational influences modulate the diffusion of independent frequencies.

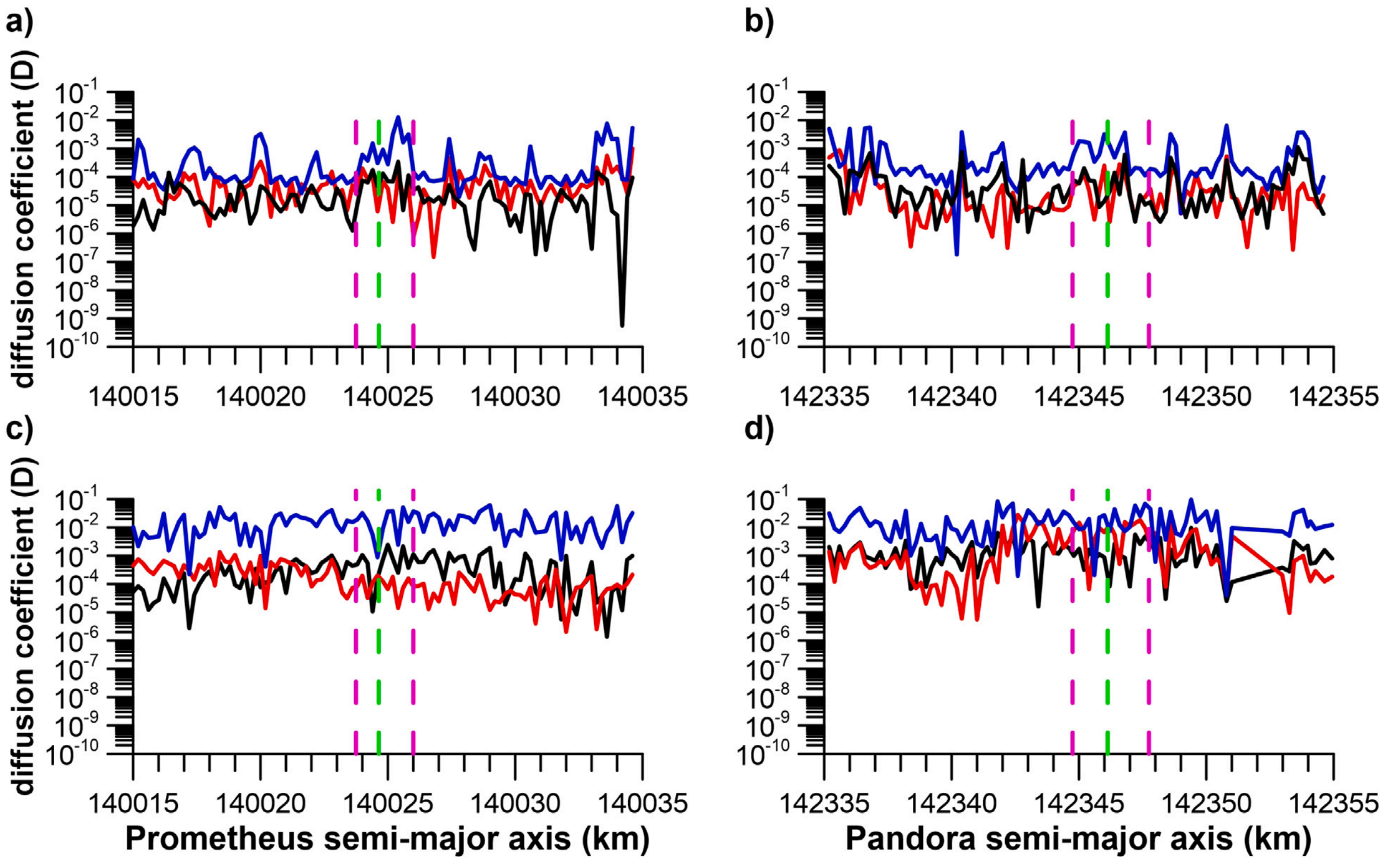

**Fig. 13.** The diffusion coefficients $\delta f_i$ are derived from the time evolution of the system's independent frequencies: the precession rate of the argument of pericentre ($f_\varpi$ — black curve), the precession rate of the longitude of the ascending node ($f_\Omega$ — red curve), and the mean motion ($n$ — blue curve). (a) and (b) illustrate the dynamical behaviour of Prometheus and Pandora, respectively, under the influence of Saturn's oblateness ($J_2$) and their mutual gravitational perturbations. Near their nominal semi-major axis values ($a_0$, marked by green dashed vertical lines), the diffusion coefficients are approximately $\delta f_i \sim 10^{-4}$, consistent with regular or quasi-periodic motion. Higher diffusion values are observed near the location of the 83:81 mean-motion resonance (see Fig. 1(a)). The magenta dashed vertical lines denote the domain of the 121:118 resonance. (c) and (d) show the results for Prometheus and Pandora, respectively, when perturbations from Mimas and Tethys are included. In this scenario, the dynamical structure is significantly altered. For Prometheus, the original resonant pattern is disrupted; although $\delta f_n$ increases, it still retains traces of the underlying resonance. At larger semi-major axis, $\delta f_\varpi$ becomes the dominant contributor to chaotic diffusion, while at smaller semi-major axis, $\delta f_\Omega$ plays a more significant role. In Pandora's case, the inclusion of additional perturbations leads to an increase in diffusion coefficients by up to two orders of magnitude in some regions. Near Pandora's nominal position, chaotic diffusion is primarily driven by variations in $\delta f_\Omega$. Despite these increases, all values of $\delta f_i$ remain below $10^{-2}$, indicating the absence of strong chaotic behaviour and the prevalence of weak or confined chaos throughout the studied domain.

For Prometheus, the simulations span the initial semi-major axis from 140,015 km to 140,035 km, covering the extension of the 121:118 resonance, with fixed eccentricity $e_0 = 0.0025$. For Pandora, the initial semi-major axis ranges from 142,335 km to 142,355 km, with $e_0 = 0.0015$. All simulations are integrated for 500 years. The resulting values $\delta f_i$ for $f_\varpi$, $f_\Omega$, and $n$ are shown in Fig. 13.

The alternation in dominance between $f_\varpi$ and $f_\Omega$ reflects their different sensitivities to external perturbations: the pericentre mode is more sensitive than the nodal mode. Changes in $f_\varpi$ are primarily associated with variations in secular dynamics related to eccentricity and resonant interactions involving the circulation of the pericentre. In contrast, the variations in $f_\Omega$ are related to the secular response associated with the inclination of the system. As the perturbations induced by Mimas and Tethys modify the pericentre and nodal precession rates through Saturn's oblateness and mutual coupling terms, different regions of phase space become sensitive to one secular mode or the other. This behaviour is consistent with the complex overlap of resonant and secular perturbations observed in the dynamical maps.

The diffusion analysis reveals predominantly regular or quasi-periodic motion near the nominal positions of Prometheus and Pandora (marked by vertical dashed lines), with $\delta f_i \sim 10^{-4}$. Localised increases in diffusion are observed near resonant separatrices, such as the 83:81 resonance at $a \approx 140,035$ km. These results are consistent with the rich resonant structure previously reported.

When perturbations from Mimas and Tethys are included, the dynamical behaviour changes significantly. For Prometheus, the original resonant features are partially disrupted: $\delta f_n$ increases by an order of magnitude, while $\delta f_\varpi$ dominates diffusion at larger semi-major axis and $\delta f_\Omega$ at smaller ones. Pandora exhibits even more significant changes: diffusion coefficients increase by up to two orders of magnitude in some regions, particularly due to variations in $\delta f_\Omega$.

Despite this enhanced diffusion, most values remain below the strong chaos threshold ($\delta f_i < 10^{-2}$), suggesting the prevalence of weak chaos in the system. The loose symmetry observed between Prometheus and Pandora in this perturbed scenario is consistent with expectations for a three-body problem, although further perturbations introduce asymmetries that are explored in later sections.

The relatively small diffusion coefficients obtained near the nominal configurations of Prometheus and Pandora suggest that the chaotic behaviour observed in the system is not linked to rapid orbital dispersion. Instead, the dynamics are characterised by confined chaos, whereby the trajectories undergo slow frequency modulation while remaining confined to narrow resonant layers in phase space. In this regime, resonant overlap generates irregular motion without producing large-scale changes to the orbital elements, thereby preserving the system's overall dynamical structure.

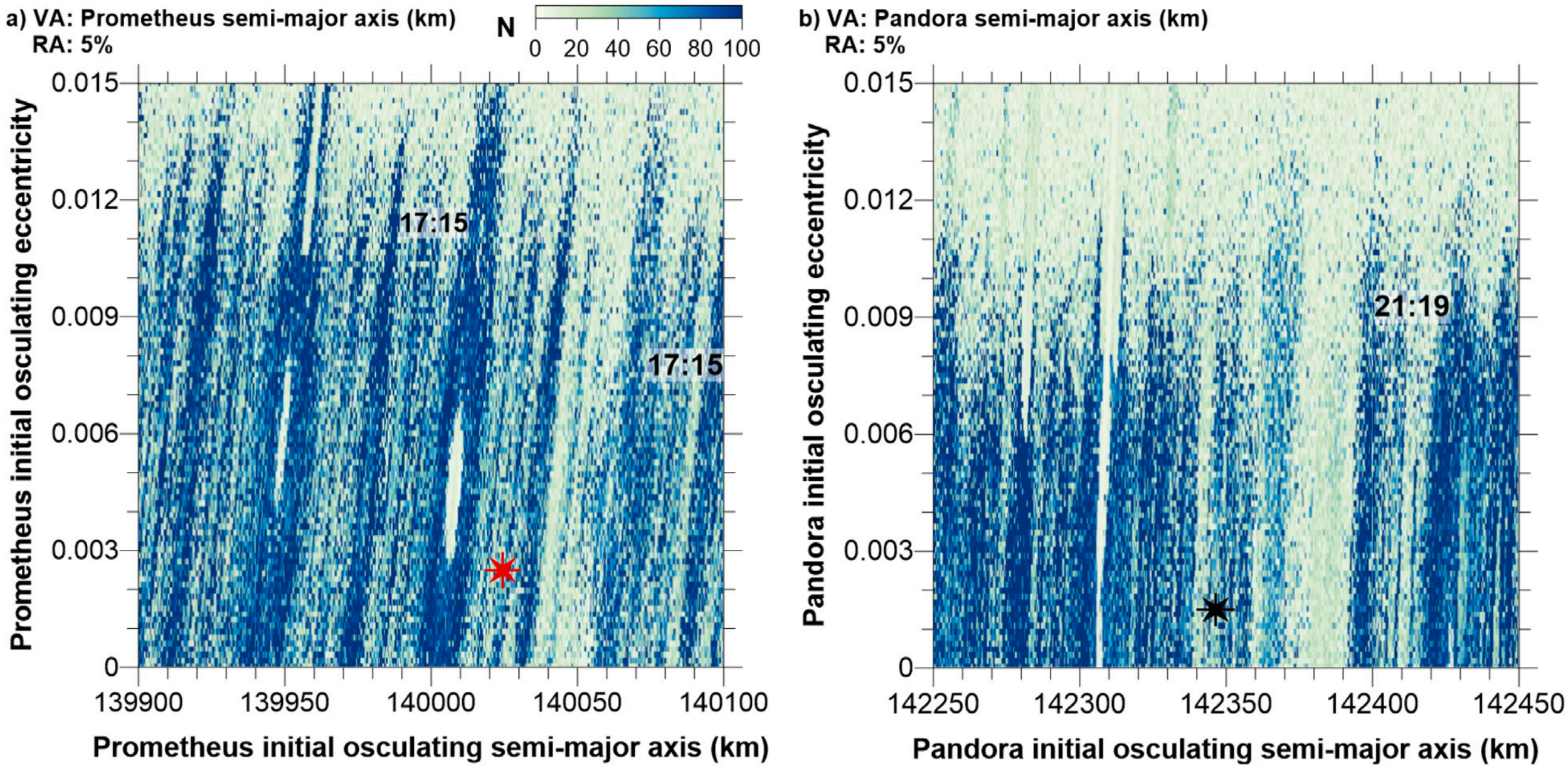

**Fig. 14.** Dynamical maps of the orbital neighbourhood of (a) Prometheus and (b) Pandora, considering perturbations from the co-orbital moons Janus and Epimetheus, as well as Mimas and Tethys. The vertical structures correspond to mean-motion resonances driven by Epimetheus, specifically the 17:15 resonance for (a) Prometheus and the 21:19 resonance for (b) Pandora. The red and black stars represent the current orbits of Prometheus and Pandora, respectively.

Therefore, the frequency-diffusion analysis indicates that the dynamics of the Prometheus–Pandora system is predominantly governed by weak chaos. Although the system exhibits resonant overlap and irregular temporal behaviour, the associated diffusion rates are relatively low, suggesting that motion is dynamically constrained rather than globally unstable. This result supports the idea that the observed chaotic behaviour mainly arises from local transport across neighbouring resonant layers rather than from large-scale stochastic evolution.

## 5. The effects of the gravitational disturbances caused by Janus and Epimetheus on Prometheus and Pandora

The Janus and Epimetheus satellites share the same orbit in a horseshoe configuration and swap their relative positions every four years. This creates a unique, cyclical, and perturbative environment for Prometheus and Pandora. Cooper and Murray (2004) demonstrated that Prometheus' longitudinal variability persists even in the absence of Pandora, provided that the Janus–Epimetheus pair is included. This indicates that co-orbitals play a significant, albeit secondary, role in the system's long-term evolution. These interactions cause the resonances induced by the moons to shift periodically. Epimetheus generates the 17:15 resonance near Prometheus and the 21:19 resonance near Pandora. Fig. 14 presents the orbital neighbourhoods of (a) Prometheus and (b) Pandora. The maps were constructed by considering the mutual perturbations between Prometheus and Pandora, as well as those from Mimas–Tethys and the Janus–Epimetheus pair.

Our dynamic maps of the orbital regions of Prometheus reveal diagonal structures that confirm the resonant proximity of these two moons. By analysing the IPS, we accurately identified the domains of these multiplets, which we have formally as $\Sigma$ multiplets.

### 5.1. Epimetheus–Prometheus

Fig. 14(a) shows Prometheus's current orbit (red star), surrounded by diagonal structures that correspond to multiplets of the 17:15 resonance with Epimetheus. Fig. 15(a) shows the IPS obtained for Prometheus's current position. These identify six main multiplets ($\Sigma_1$–$\Sigma_6$), each of which is associated with a specific critical angle. Although $\sigma_4$ and $\sigma_5$ exhibit circulation, $\sigma_1$, $\sigma_2$, $\sigma_3$, and $\sigma_6$ librate around 180°, revealing distinct dynamical behaviours within the resonance; these angles are illustrated in Figs. 15(b–g), and the corresponding semi-major axis values for each Prometheus clone are listed in Table 3.

**Table 3**
Multiplets associated with the 17:15 resonance with Epimetheus. The parameter $a_0$ denotes the initial semi-major axis of each Prometheus clone. The critical angle for each multiplet is explicitly defined, where the subscripts $Epi$ refer to Epimetheus.

| Multiplet | $a_0$ (km) | Critical angle |
|---|---|---|
| $\Sigma_4$ | 139,983.0 | $\sigma_4 = 17\lambda_{\rm Epi} - 15\lambda_{\rm cPro} - 2\Omega_{\rm cPro}$ |
| $\Sigma_5$ | 139,990.3 | $\sigma_5 = 17\lambda_{\rm Epi} - 15\lambda_{\rm cPro} - \Omega_{\rm Epi} - \Omega_{\rm cPro}$ |
| $\Sigma_6$ | 139,997.6 | $\sigma_6 = 17\lambda_{\rm Epi} - 15\lambda_{\rm cPro} - 2\Omega_{\rm Epi}$ |
| $\Sigma_3$ | 140,083.0 | $\sigma_3 = 17\lambda_{\rm Epi} - 15\lambda_{\rm cPro} - 2\varpi_{\rm Epi}$ |
| $\Sigma_2$ | 140,090.1 | $\sigma_2 = 17\lambda_{\rm Epi} - 15\lambda_{\rm cPro} - \varpi_{\rm Epi} - \varpi_{\rm cPro}$ |
| $\Sigma_1$ | 140,097.25 | $\sigma_1 = 17\lambda_{\rm Epi} - 15\lambda_{\rm cPro} - 2\varpi_{\rm cPro}$ |

*Note:* The eccentricity is fixed at $e_0 = 0.0025$ for each Prometheus clone.

**Table 4**
Multiplets and critical angle expressions associated with the 21:19 resonance with Epimetheus. The subscript $cPand$ denotes the Pandora clone.

| Multiplet | $a_0$ (km) | Critical angle |
|---|---|---|
| $\Sigma_3$ | 142,404.9 | $\sigma_3 = 21\lambda_{\rm Epi} - 19\lambda_{\rm cPand} - 2\varpi_{\rm Epi}$ |
| $\Sigma_2$ | 142,409.0 | $\sigma_2 = 21\lambda_{\rm Epi} - 19\lambda_{\rm cPand} - \varpi_{\rm Epi} - \varpi_{\rm cPand}$ |
| $\Sigma_1$ | 142,414.0 | $\sigma_1 = 21\lambda_{\rm Epi} - 19\lambda_{\rm cPand} - 2\varpi_{\rm cPand}$ |

*Note:* The eccentricity is fixed at $e_0 = 0.0015$ (IPS) for each Pandora clone.

### 5.2. Epimetheus–Pandora

Fig. 14(b) shows the current orbit of Pandora (black star), which is located approximately 60 km away from the 21:19 resonance domain with Epimetheus. Despite this distance, the orbital mapping reveals clear vertical structures in the neighbourhood of Pandora. Three multiplets ($\Sigma_1$, $\Sigma_2$, $\Sigma_3$) identified just beyond Pandora's current orbit are

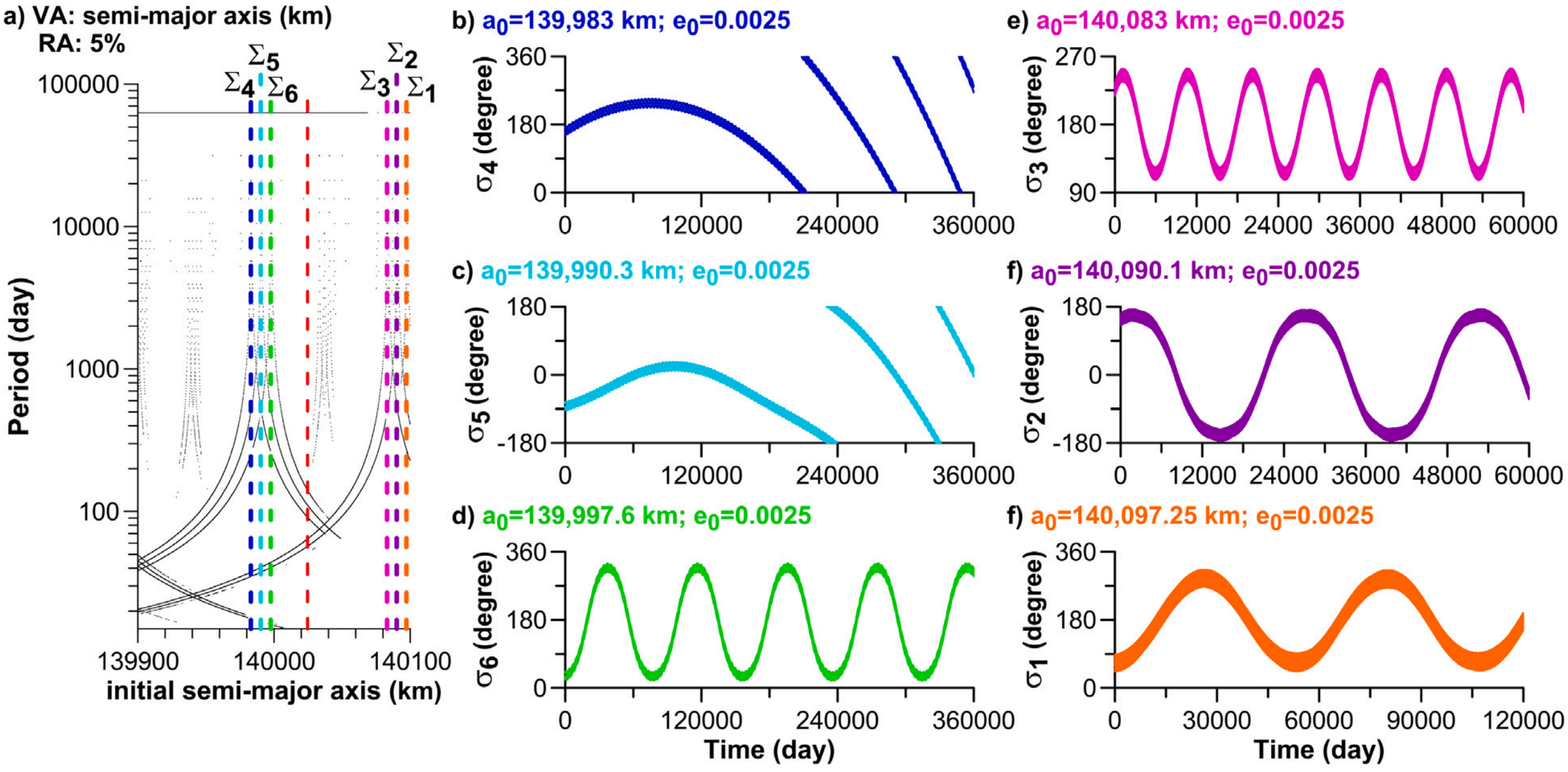

**Fig. 15.** (a) IPS for the Prometheus semi-major axis, with RA = 5% and considering Janus, Epimetheus, Mimas and Tethys as the main perturbing bodies. Coloured dotted lines represent the initial semi-major axis ($a_0$) for Prometheus clones within the multiplets associated with the 17:15 resonance with Epimetheus; the red dashed line indicates the current semi-major axis of Prometheus. (b–f) display the temporal variations of the critical angles for each multiplet identified in (a). Circulation of the critical angles is observed in (b) and (c), whereas (d–f) exhibit oscillations around 180°. Specific details for each multiplet, including semi-major axis values and critical angle descriptions, are provided in Table 3.

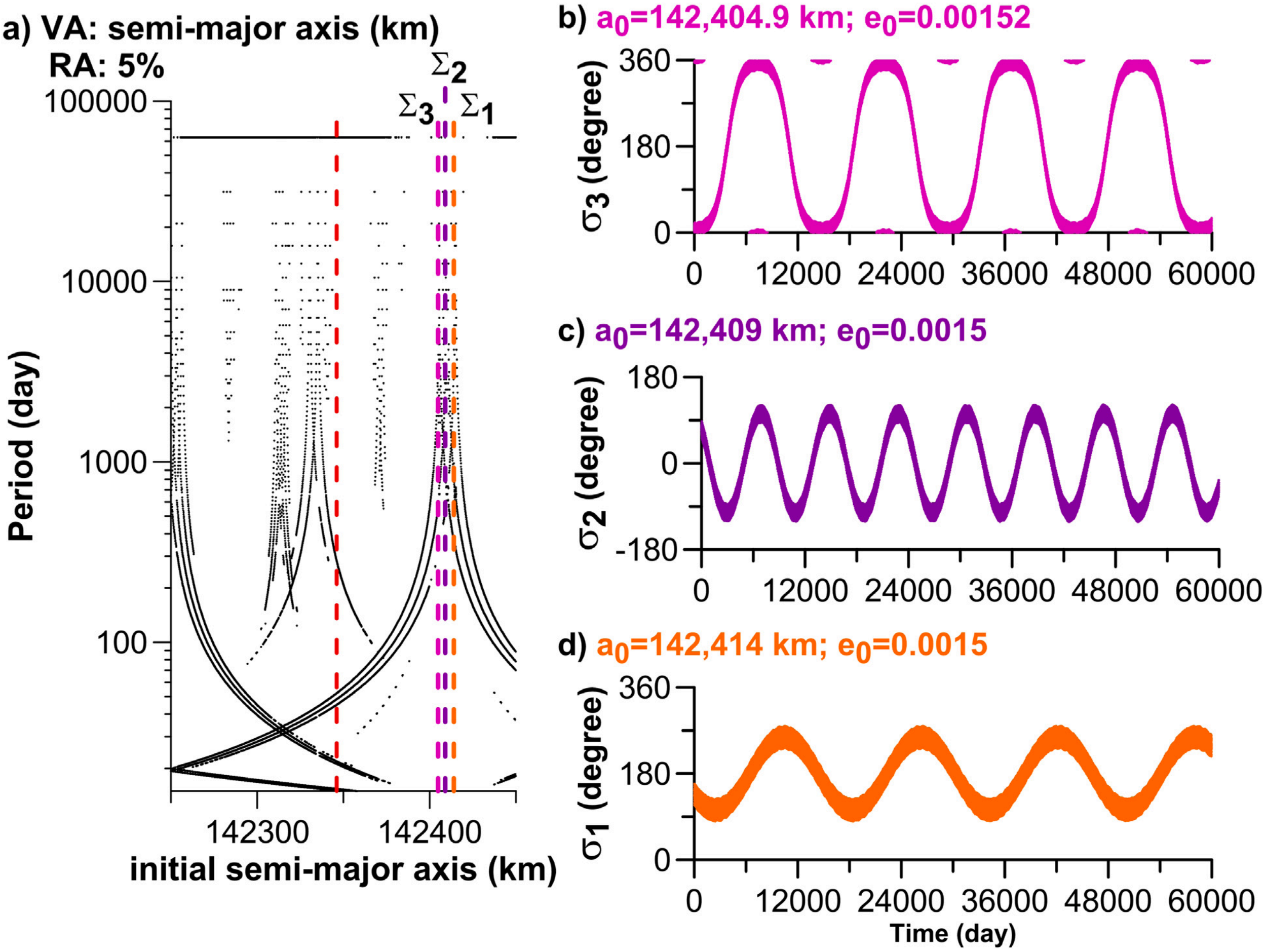

**Fig. 16.** (a) IPS for the semi-major axis of Pandora, with RA = 5% and considering Janus, Epimetheus, Mimas and Tethys as the main perturbing bodies. Coloured dotted lines indicate the initial semi-major axis ($a_0$) for Pandora clones within the multiplets associated with the 21:19 resonance with Epimetheus; the red dashed line represents Pandora's current semi-major axis. (b–d) display the temporal evolution of the critical angles for each multiplet identified in (a). Circulation is observed in (b), while (c) and (d) exhibit libration around 180°.

**Table 5**
Parameter values for Saturn's core and envelope.

| Parameter | Symbol | Value |
|---|---|---|
| Mass of Saturn's core | $M_c$ (kg) | $18.65M_\oplus$ |
| Radius of Saturn's core | $R_c$ (km) | $0.219R_S$ |
| Saturn core density | $\rho_c$ (kg/m$^3$) | 1,156.3 |
| Density of Saturn's envelope | $\rho_0$ (kg/m$^3$) | 503.587 |
| Rigidity of Saturn's core | $\eta_c$ | $10^{14}$ Pa s |
| Viscosity of Saturn's core | $\mu_C$ | $10^{11}$ Pa |
| Relaxation factor | $\gamma$ | $10^{-11}$ s$^{-1}$ |
| Rotation rate of Saturn | $\Omega$ (rad s$^{-1}$) | $1.64\times10^{-4}$ |

*Note:* The Earth's mass is given by $M_\oplus = 5.9736\times10^{24}$ kg (Remus et al., 2015; Shoji and Hussmann, 2017) and $R_S$ represents the equatorial radius of Saturn given in Table 1.

illustrated in Figs. 16(b–d). Of these, the critical angle $\sigma_1$ exhibits circulation, while $\sigma_1$ and $\sigma_2$ librate around 180°, demonstrating different dynamic behaviours within the resonance. The corresponding semi-major axis values for each Pandora clone are listed in Table 4.

## 6. The dissipation and migration of Pandora: The recent past of Prometheus

Both Saturn and Pandora experience dissipative tidal effects that alter the satellite's semi-major axis and eccentricity. As we are working with a gas giant and an icy moon, the modelling of tidal evolution differs for each body. For Saturn, we will consider a viscous core surrounded by a gaseous layer. The dissipative effects in each layer are described by *Maxwell's rheology* (Shoji and Hussmann, 2017; Remus et al., 2015) and *resonant locking (RL)* as described by Witte and Savonije (1999, 2001). The dissipative effects on Pandora will be analysed based on the *creep model* described in Ferraz-Mello (2013, 2015), which assumes that the moon is a homogeneous and viscous body in synchronous rotation with Saturn. Following Ceccatto et al. (2025) and Rodríguez et al. (2025), this section determines Pandora's migration driven by tidal dissipative effects. The physical parameters for solving the tidal equations described in the subsections below are summarised in Table 5.

### 6.1. Tidal dissipation in Saturn

Saturn is modelled as a two-layer body, consisting of a viscoelastic core surrounded by a gaseous envelope. Each layer contributes separately to tidal dissipation. According to Ferraz-Mello et al. (2008), the equations describing the tidal effects within Saturn, based on *Darwin's model*, are presented as follows:

$$\dot{a}_{Saturn} = \frac{2a}{3n}\frac{9n^2 M_{Pand} R_S^5}{2M_S a^5}\frac{k_2}{Q}(1+\frac{51}{4}e^2), \tag{9}$$

$$\dot{e}_{Saturn} = \frac{57neM_{Pand}R_S^5}{8M_S a^5}\frac{k_2}{Q}, \tag{10}$$

where $M_{Pand}$ is the mass of Pandora, $M_S$, and $R_S$ are the mass and radius of Saturn, $a$, $e$, and $n$ are the semi-major axis, eccentricity, and mean angular velocity of Pandora, respectively. The ratio $k_2/Q$ is given by the second-order Love number $k_2$ and the dissipation function $Q$. The following subsections describe tidal dissipation in both the core and the gas envelope.

#### 6.1.1. Dissipation in the Saturn's core

Maxwell's rheology describes the energy dissipation in the planet's viscoelastic core. This is considered through the equations in Shoji and Hussmann (2017) and Remus et al. (2015)

$$\widetilde{k_{2s}} = \frac{3}{2}\frac{\widetilde{\epsilon}+\frac{2}{3}\beta}{\alpha\,\widetilde{\epsilon}-\beta}, \tag{11}$$

$$\alpha = 1+\left(\frac{5}{2}\right)\left(\left(\frac{\rho_c}{\rho_0}\right)-1\right)\left(\frac{R_c}{R_S}\right)^3, \tag{12}$$

$$\beta = \frac{3}{5}\left(\frac{R_c}{R_S}\right)^2(\alpha-1), \tag{13}$$

$$\widetilde{\epsilon} = \frac{\frac{19\widetilde{\mu}_c}{2\rho_c g_c R_c}+\frac{\rho_0}{\rho_c}\left(1-\frac{\rho_0}{\rho_c}\right)\left(\beta+\frac{3}{2}\right)+\left(1-\frac{\rho_0}{\rho_c}\right)}{\left(\alpha+\frac{3}{2}\right)\frac{\rho_0}{\rho_c}\left(1-\frac{\rho_0}{\rho_c}\right)}, \tag{14}$$

being $Q = |\widetilde{k_{2s}}|/|\mathrm{Im}[\widetilde{k_{2s}}]|$, $\mathrm{Im}[\widetilde{k_{2s}}] = k_{2s}/Q$, with $k_2 = \widetilde{k_{2s}}$. The parameters $g_c$, $R_c$, $\rho_c$ are the gravitational acceleration on the surface of the core, the radius of the core, and its density. $\rho_0$ is the density of the envelope, and the constants $\alpha$ and $\beta$ are depends on the physical parameters of the planet. The complex of Maxwell's frequency Remus et al. (2015) is given by

$$\tilde{\mu}_c = \frac{w_f\mu_c\eta_c}{w_f\eta_c+i\mu_c}, \tag{15}$$

where $i=\sqrt{-1}$, $\mu_c$ and $\eta_c$ are the rigidity and viscosity of the core, the $w_f = 2(\Omega-n)$ is the frequency associated with the planet's rotation speed, $\Omega$. Table 5 presents values for Saturn's core and envelope parameters. The other parameters were calculated.

#### 6.1.2. Dissipation in the Saturn's gas envelope

Planets such as Saturn have convective envelopes in which the oscillating fluid generates inertial waves. These inertial waves are considered attractors for tidal energy dissipation (Fuller et al., 2016). As the planet's internal structure evolves, a satellite can become locked near a resonance and migrate. This is the RL mechanism. To describe this mechanism, we use the work of Lainey et al. (2020) where the dissipative ratio is given

$$\frac{k_2}{Q} = \frac{B}{3}\frac{M_S^{1/2}}{G^{1/2}M_{Pand}R_S^5}\frac{a_c^{1/B}}{t_c}a^{-1/B+13/2}, \tag{16}$$

where $M_S$ and $R_S$ are the mass and radius of Saturn, and $M_{Pand}$ is the mass of Pandora. The parameters $a_c$ and $t_c$ are Pandora's current semi-major axis and Saturn's age, while $B$ is a constant associated with the motion of the gas in Saturn's envelope.

### 6.2. Tidal dissipation on Pandora

To describe the tidal dissipation on Pandora, we use creep theory, which is similar to Maxwell's model in rheology due to viscosity being the fundamental parameter in both (Ferraz-Mello, 2013, 2015). It is worth mentioning that we model Pandora as a homogeneous satellite that is in synchronous rotation with Saturn. Considering Folonier and Ferraz-Mello (2017) we have

$$\gamma = \frac{3gM_{Pand}}{8\pi R_S^2\eta_P}, \tag{17}$$

$$\epsilon_\rho = \frac{15}{4}\frac{M_{Pand}}{M_S}\left(\frac{R_{Pand}}{a}\right)^3, \tag{18}$$

and

$$\dot{a}_{Pand} = -\frac{42R_S^2\bar{\epsilon}_\rho e^2}{5a}\frac{n^2\gamma}{n^2+\gamma^2}, \tag{19}$$

$$\dot{e}_{Pand} = -\frac{21R_S^2\bar{\epsilon}_\rho e\left(1-e^2\right)}{5a^2}\frac{n^2\gamma}{n^2+\gamma^2}. \tag{20}$$

The parameter $\gamma$ is the relaxation factor and is inversely proportional to the viscosity of the satellite. For Pandora, we assume $\gamma = 10^{-11}$ s$^{-1}$.

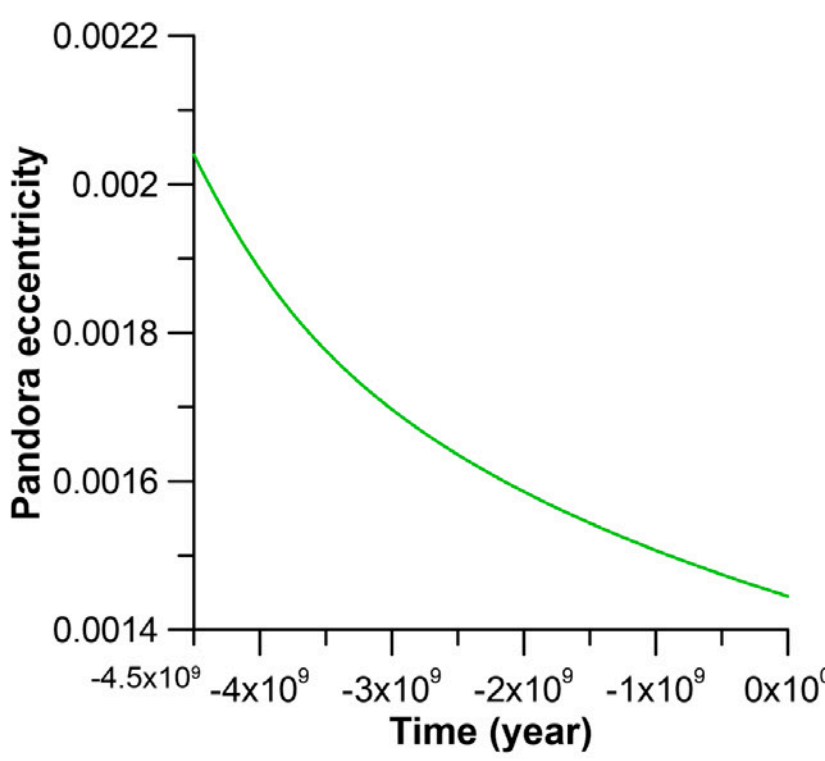

**Fig. 17.** Variation in (a) semi-major axis and (b) eccentricity of Pandora is considered, taking into account the viscosity of the core $\eta_c = 10^{14}$ Pa.s and the relaxation factor $\gamma = 10^{-11}$ s$^{-1}$ Pa.s, allowing for a slow variation in the eccentricity of Pandora. The values for the current semi-major axis of Prometheus and Pandora are represented by the red and blue horizontal lines respectively. We estimate that Pandora was close to Prometheus at about $-2.12 \times 10^8$ years ago.

### 6.3. Temporal variation of orbital parameters

Adopting the physical parameters listed in Table 5, a simple order-of-magnitude analysis supports the hierarchy between the tidal contributions discussed above.

Considering the orbital elements of Pandora ($a = 1.42364 \times 10^8$ m and $e = 1.5 \times 10^{-3}$), together with $R_{Pand} \sim 4.0 \times 10^4$ m, $M_{Pand} \sim 1.4 \times 10^{17}$ kg, $M_S = 5.68 \times 10^{26}$ kg, $R_S = 6.03 \times 10^7$ m, and $n \sim 1.1 \times 10^{-4}$ s$^{-1}$, and adopting a representative value $k_2/Q \sim 10^{-4}$ for Saturn, consistent with the viscoelastic and resonance locking models described above, Eqs. (9) and (19) yield

$(\dot{a})_{\rm Saturn} \sim 10^{-11}$ m s$^{-1}$.

In contrast, Eqs. (19), (17) and (18), assuming $\gamma \sim 10^{-11}$ s$^{-1}$, give the following

$(\dot{a})_{\rm Pandora} \sim 10^{-13}$ m s$^{-1}$,

implying that the tidal expansion driven by Saturn exceeds the internal tidal contribution of Pandora by roughly two to three orders of magnitude. This strong asymmetry results from the scaling $(R_S/a)^5$ combined with the large planetary radius, whereas the satellite contribution is further suppressed by the small eccentricity through the factor $e^2$.

A similar hierarchy is obtained for the eccentricity evolution. Eq. (10) gives the following

$(\dot{e})_{\rm Saturn} \sim 10^{-21}$ s$^{-1}$,

while Eq. (20) produces

$(\dot{e})_{\rm Pandora} \sim 10^{-12}$ s$^{-1}$.

Therefore, the eccentricity damping associated with internal dissipation within Pandora overwhelmingly dominates the corresponding Saturnian contribution. These estimates indicate a clear separation between the dominant tidal mechanisms controlling the long-term evolution of the system: the semi-major axis expansion is primarily governed by tides raised on Saturn, whereas the eccentricity evolution is controlled mainly by internal dissipation within Pandora. Motivated by this hierarchy, we numerically integrate the coupled system formed by Eqs. (9) and (20).

To construct Fig. 17, we used the dissipative ratios[2] referring to each layer in Equation (9) integrating it with Equation (20) over a period of $4.5 \times 10^9$ years. We assumed the following values for the parameters: $\eta_c = 10^{14}$ Pa.s, $\mu_c = 10^{11}$ Pa, $\gamma = 10^{-11}$ s$^{-1}$ and $B = 1/3$.

Rossignoli et al. (2019) estimated the surface ages of Saturn's small satellites based on crater counts. They determined that the age of Prometheus was approximately $\tau_{\rm Pro} \approx 1 \times 10^9$ years and that of Pandora around $\tau_{\rm Pan} \approx 2 \times 10^9$ years.

### 6.4. Revisiting the past: the dynamical map

The effects of Pandora's tidal expansion on the orbital dynamics of Prometheus were analysed using dynamical mapping by varying Pandora's semi-major axis and eccentricity. This mapping was presented by Giuppone et al. (2022) when investigating the effect of Charon's orbital expansion on Pluto's other six small satellites and was subsequently applied by Ceccatto et al. (2025) to study the dynamics of Atlas during Prometheus's tidal expansion.

We constructed a grid of numerical simulations, considering Saturn as the main body and analysing the expansion of Pandora. We also accounted for Saturn's oblateness by including the $J_2$, $J_4$ and $J_6$ terms of its gravitational potential.

Using the MERCURY package (Chambers, 1999) and the RA 15 algorithm (Everhart, 1985), we numerically solved the full motion equation for each initial condition present in the grid. Following (Rodríguez and Callegari, 2021), we calculate the maximum variation in the semi-major axis of Prometheus ($\Delta a$). The initial orbital elements for Prometheus and Pandora are provided in Table 2, except for the semi-major axis and eccentricity for Pandora, which were varied to generate the grid used in the dynamical map.

From Fig. 17(a), we can see that Pandora reached Prometheus at an approximate age of $\tau_{\rm Pand} \approx 2.12 \times 10^8$ years (red lines dashed). Using this age as a reference point, we can construct our dynamical map. We take a semi-major axis value of 139,900 km for Pandora as our starting point, which corresponds to an age of $\tau_{\rm Pand} \approx 2.24 \times 10^8$ years.

We built the map shown in Fig. 18 by adopting [139,900 km, 142,400 km] for the semi-major axis and [0, 0.015] for the eccentricity of Pandora clones. The black and blue stars represent the current orbital positions of Prometheus and Pandora, respectively. The dotted green line shows Pandora's tidal migration. High values of $\log \Delta a$ indicate irregular motion, while low values suggest regular motion. The magenta line represents the classic collision curve[3] between Pandora and Prometheus during Pandora's tidal migration. This curve describes the trajectory in which the minimum distance between the orbits of the two bodies' orbits approaches the critical distance required for collision.

[2] Calculating the dissipative ratio, $\frac{k_2}{Q}$, due to each layer, we find that for the core, $\left[\frac{k_2}{Q}\right]_{\rm core} \approx 3.38 \times 10^{-5}$ and for the gas envelope, $\left[\frac{k_2}{Q}\right]_{\rm envelope} \approx 1.7 \times 10^{-3}$. These results indicate that the influence of dissipation in the gas envelope dominates the variation of Pandora's semi-major axis.

[3] The curve is defined by the relationship between the semi-major axis and eccentricity of both bodies: $a_{\rm Pro}(1 + e_{\rm Pro}) = a_{\rm cPand}(1 - e_{\rm cPand})$.

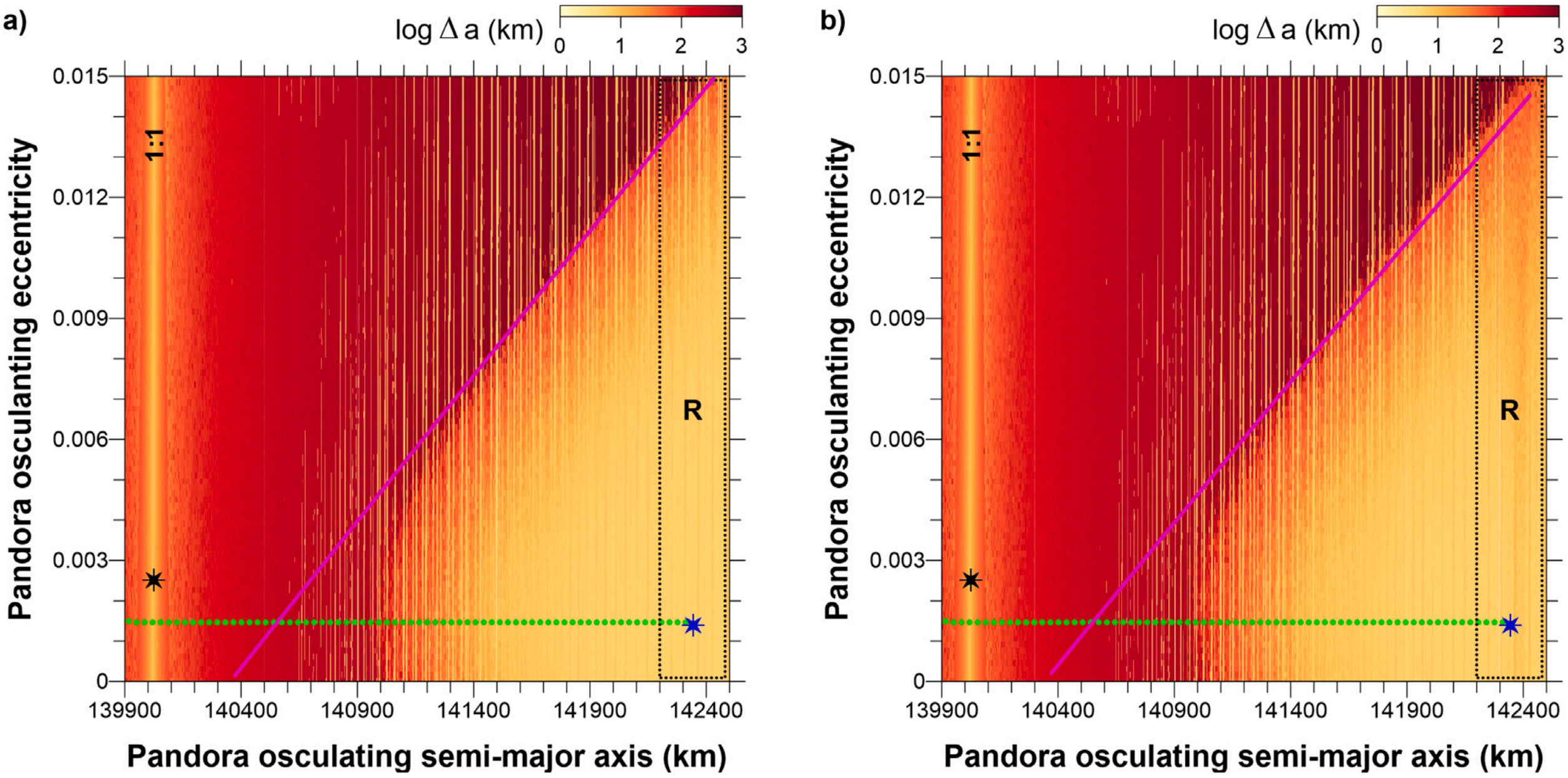

**Fig. 18.** (a) Dynamical map for Prometheus shows how the initial semi-major axis and eccentricity of Pandora vary. The model includes Prometheus and Pandora, as well as Saturn's potential coefficients $J_2$, $J_4$, and $J_6$. The initial orbital elements of Prometheus were fixed on 1 January 2000. We performed 132,613 numerical simulations, integrating over 100 years with a time step of 0.1 years. For each simulation, we calculated the maximum variation in Prometheus's semi-major axis ($\Delta a$). The red and blue stars represent the current orbital positions of Prometheus (140,024.64 km, 0.00252) and Pandora (142,346.13 km, 0.0015), respectively. Regions with high $\log \Delta a$ values correspond to irregular motion, while regions with the low $\log \Delta a$ values represent resonance locations between Prometheus and Pandora. The green dotted line illustrates Pandora's tidal orbital evolution. Section 6.5 discussed the region for co-orbital motion, while Section 6.6 addresses the region for first-and higher-order resonances (region **R**). The magenta curve illustrates the classical collision curve, defined by the equality between Pandora's pericentric distance and Prometheus's apocentric distance throughout Pandora's tidal migration (b) Like (a), we also include the Mimas–Tethys pair. We note that the Mimas–Tethys perturbations do not significantly alter the structures in (a).

Fig. 18(a) shows that around the current position of Prometheus, there is a vertical band approximately 140 km, wide, bordered by areas with high values of $\log \Delta a$. Within this band, there is a narrow region with low values of $\log \Delta a$, representing the co-orbital resonance zone, which is discussed in detail in Section 6.2. Additionally, Fig. 18 shows another region with low values of $\log \Delta a$ labelled **R** and delimited by a dotted rectangle near the current orbit of Prometheus. This region is discussed in Section 6.3. Fig. 18(b) shows the same mapping as Fig. 18(a) but with the Mimas–Tethys pair included. Comparing Figs. 18(a) and (b), it can be seen that the perturbative effects of the Mimas–Tethys pair do not significantly alter the resonant structures observed in Fig. 18(a).

### 6.5. Revisiting the past: the co-orbital scenario

In the context of the problem of three bodies, co-orbital motion (or 1:1 mean-motion resonance) exhibits two distinct regimes of libration for the critical angle, $\Delta\lambda$, which is defined as the difference in mean longitude between the orbiting bodies. In the first regime, the critical angle $\Delta\lambda$ librates around the Lagrange equilibrium points, $L_4 = 60°$ or $L_5 = 300°$, characterising the "tadpole" orbit regime. In the second regime, the critical angle $\Delta\lambda$ librates with a large amplitude around 180°, also encompassing the $L_4$ and $L_5$ points. This is known as the "horseshoe" orbit regime.

Fig. 19(a) shows the phase-space mapping for the 1:1 mean-motion resonance within the semi-major axis interval [139,940 km, 140,100 km]. The inner diagonal white line represents the equilibrium solution, and the black star indicates Prometheus's current orbit. The green dotted curve illustrates Pandora's tidal evolution. This scenario illustrates the co-orbital motion of the Prometheus-"Pandora clone" pair. The boundaries of this region correspond to high values of $\log \Delta a$, which are associated with separatrices.

Figs. 19(b–e) show how the geometric angle $\Delta\lambda$ varies over time for four initial conditions, as indicated by the coloured symbols in Fig. 19(a). The cyan circles in Fig. 19(a) correspond to initial conditions where the orbit follows the tadpole regime (Figs. 19(b), and (c)), while the blue triangles correspond to initial conditions with horseshoe orbits (Figs. 19(d), and (e)).

### 6.6. Revisiting the past: the orbital neighbourhood near Pandora

The initial semi-major axis of 142,200 in Fig. 17(a) corresponds to a Pandora age equivalent to $\tau_{\mathrm{Pand}} \approx 3.19 \times 10^7$ years, with an eccentricity value of ≈ 0.00144. Fig. 20(a) illustrates the orbital neighbourhood of Pandora (blue star), which corresponds to region **R** in Fig. 18(a). Several crossing resonances were observed in the past (vertical strips), with first-order resonances being the strongest, albeit with a low probability of capture (see Table 6). The 121:118 resonance appears as a narrow diagonal band, within which Pandora is located. Fig. 20(b) shows the same region, now considering the perturbative effects of Mimas–Tethys. Comparing Figs. 20(a) and (b), we observe small changes in several resonant structures, in particular, the 42:41 and 40:39 resonances are significantly affected. Their internal regions are filled with separatrices, as can be seen in Fig. 1(c). The 121:118 resonance has also changed, becoming slightly wider and exhibiting a notable increase in the $\log \Delta a$ value, resulting from the expansion of its separatrices.

The magenta line represents the collision curve between Pandora and Prometheus. This curve is far from close to the green dotted line, which represents Pandora's migration due to tidal effects.

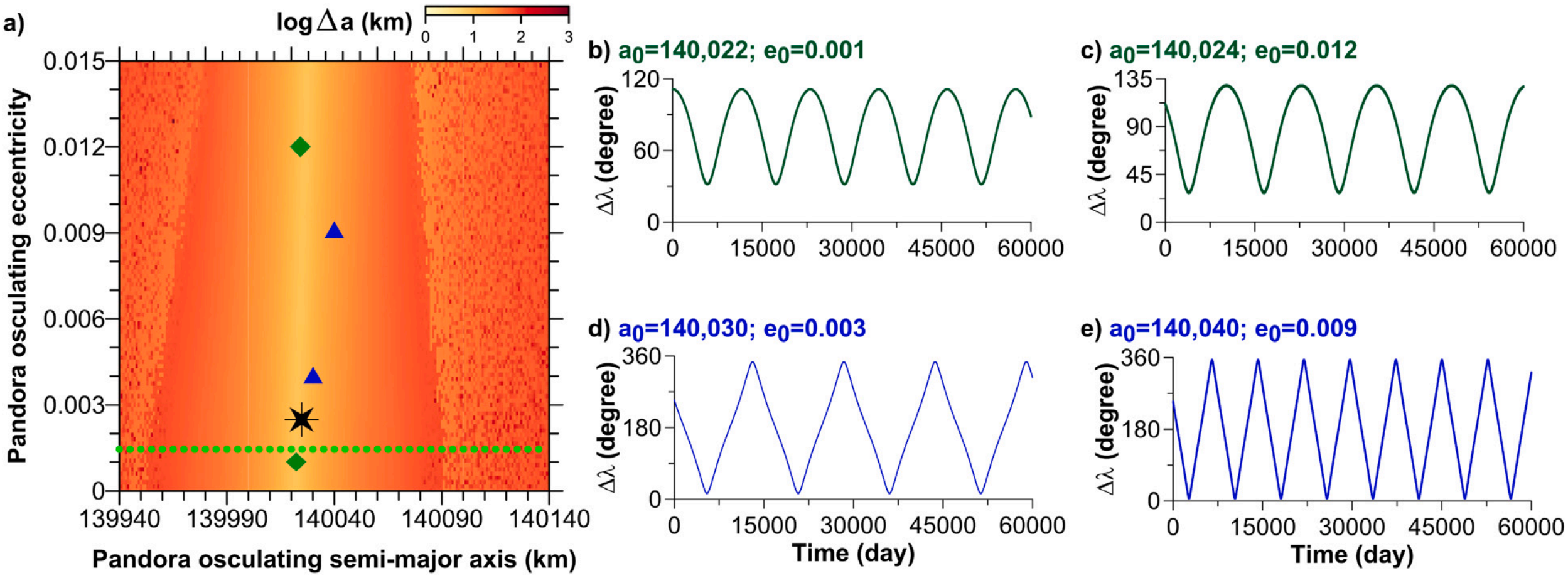

**Fig. 19.** Dynamical map for co-orbital motion. The black star indicates Prometheus's current orbit, while the coloured symbols represents Pandora clones' initial conditions. The green dotted curve illustrates Pandora's tidal evolution. (a) shows the co-orbital resonance zone, with the lighter band indicating the equilibrium solutions. (b) and (c) show examples of tadpole orbital motion, and (d) and (e) depict horseshoe orbital motion. A total of 20,402 numerical simulations were performed in (a), integrated over 100 years with a time step of 0.1 days.

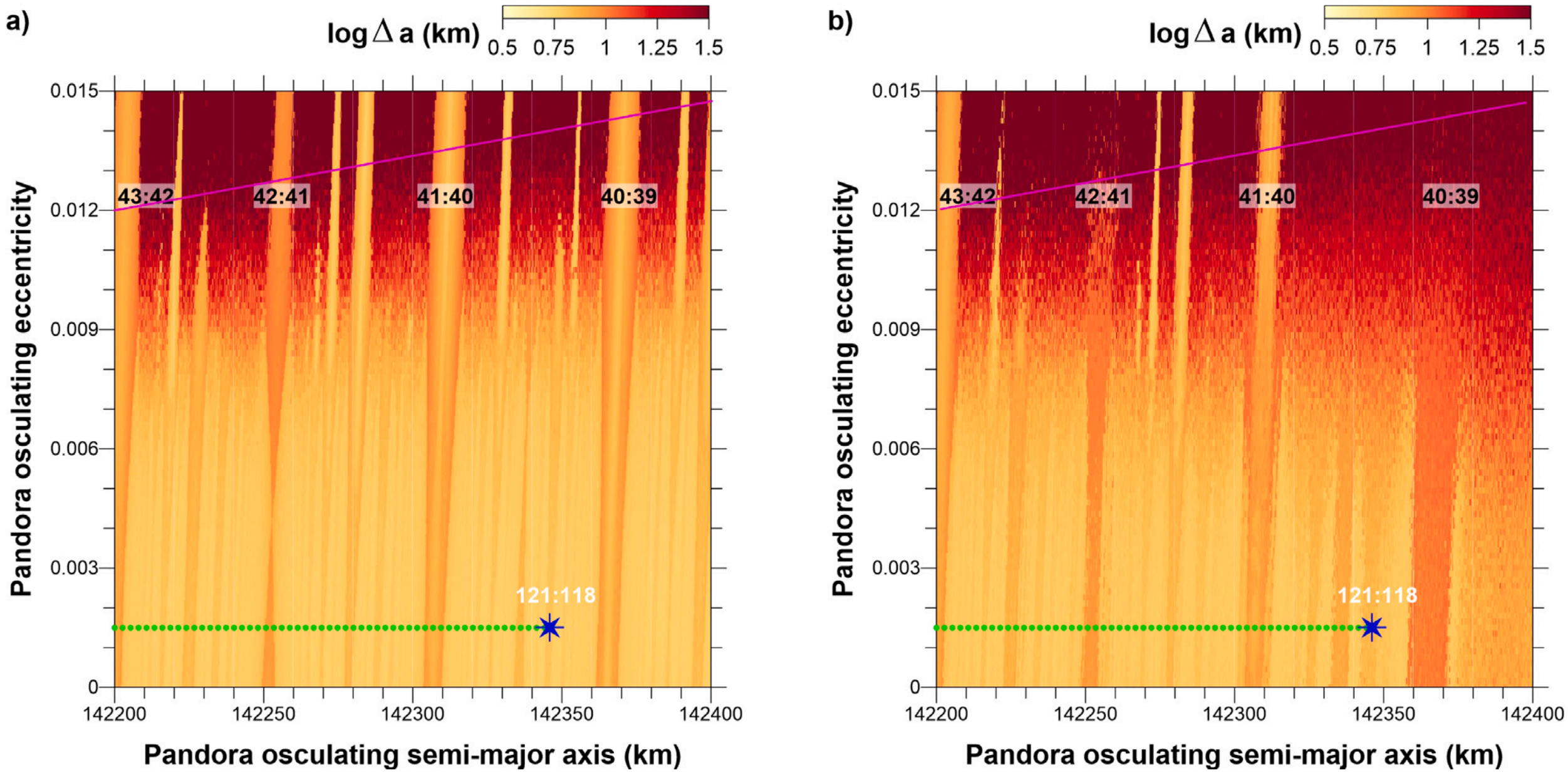

**Fig. 20.** (a) As in Fig. 18, the **R** region representing the neighbourhood of Pandora's current orbit, indicated by the blue star, has been considered. The green dotted curve shows Pandora's tidal orbital evolution, and the magenta curve shows the classical collision curve between Prometheus and Pandora. Several resonances are crossed during Pandora's migration, with the first-order resonances being the strongest. The 121:118 resonance appears as a narrow diagonal band, with Pandora (blue star) located inside this region.(b) As in (a), but including the Mimas–Tethys pair. Including this pair results in a small change to some resonant structures, with the 42:41 and 40:39 resonances being the most affected. The width of the 121:118 resonance increased, as did its separatrices.

### *6.6.1. Capture probability*

The *capture probability* is the probability that a migrating particle or satellite will become trapped in a resonant libration region during its migration. In the context, El Moutamid et al. (2017) derived analytical expressions for the probability of capture into first-order mean-motion resonances,

$$P_m \approx \frac{W_{\mathrm{CER}}}{\pi a_0},$$

where the resonance width $W_{\mathrm{CER}}$ is given by

$$W_{\mathrm{CER}} = \frac{8a_0\sqrt{|\epsilon_c|}}{3|m|},$$

where, $a_0$ represents the current semi-major axis of Prometheus, while $m$ denotes the resonance order. The parameter $\epsilon_c$ is defined as

$$\epsilon_c = 3m^2\left(\frac{m_s}{M_c}\right)\left(\frac{a_0}{a_s}\right)A'_m e_s,$$

**Table 6**
The probabilities of capture and the widths of the corresponding $W_{\text{CER}}$ corotation mean-motion resonances shown in Fig. 20.

| Resonance | $a_S$ (km) | $W_{CER}$ (km) | Probability (%) |
|---|---|---|---|
| 43:42 | 142,205 | 2.288 | 0.00052 |
| 42:41 | 142,252.5 | 2.261 | 0.000514 |
| 41:40 | 142,307.5 | 2.233 | 0.000507 |
| 40:39 | 142,365 | 2.206 | 0.000501 |

The values $a_0$=140,024.64 km and $e_S$=0.0015 were used to determine the commensurability.

where $m_s$ and $M_c$ are the masses of the perturbing satellite (Pandora) and Saturn, respectively, $a_s$ is the semi-major axis of the perturbing satellite, $e_s$ is its eccentricity, and $A'_m \sim 0.8m$ for large values of $m$.

Table 4 shows the $W_{\text{CER}}$ and capture probability for the resonances depicted in Fig. 20.

## 7. Conclusions

In this study, we used frequency analysis primarily to investigate the orbital dynamics of the Prometheus–Pandora system. By mapping the phase space in Prometheus' orbital neighbourhood (see Fig. 1(a)), we observed several diagonal structures representing resonant domains. Among these, we highlight the 39:38, 40:39, and 41:40 resonances, as well as the associated Lindblad and corotation domains. Prometheus is in resonance with Pandora at 121:118. However, the presence of a variety of separatrices in this region made it impossible to clearly identify the domains of the multiplets $\Psi_1, \Psi_2, \Psi_3$ and $\Psi_4$ associated with this resonance.

The orbital neighbourhood of Pandora exhibited the same dynamic characteristics (see Fig. 1(b)). Within this region, we identified several high-order resonances, specifically the 82:80 and 80:78 resonances. The irregular time evolution observed in critical angles $\psi_1$, $\psi_2$. $\psi_3$, and $\psi_4$ (Fig. 4) This is consequently attributed to the shared location of Prometheus and Pandora within the separatrices of the 121:118 resonance.

In addition, the orbital elements of Prometheus, such as its semi-major axis, eccentricity, and inclination, which are affected by Pandora, revealed several periods exhibiting the characteristics of irregular, or chaotic, orbits. This phenomenon was also observed in Pandora's orbital elements, further highlighting the complexity of the system's dynamics. The overlap observed in these domains in the IPS (see Figs. 7(a) and 8(a)) suggests chaotic motion for both Prometheus and Pandora, which reinforces the dynamic complexity of the system.

Including the perturbative effects of the Mimas–Tethys pair had a direct impact on the results observed in our analysis. Comparing Figs. 1(a) and (c), as well as Figs. 1(b) and (d), reveals the perturbation of the higher-order resonant structures, resulting in significant deformation of the domains associated with the Lindblad and corotation resonances. However, the 80:78 resonance near Pandora remains unchanged. Furthermore, analysis of Figs. 3(c) and (d), which provide a detailed view of the orbital region of both moons, shows that the resonant separatrices seen in Figs. 3(a) and (b) have disappeared. These results suggest that Mimas and Tethys play a significant role in dissipating pre-existing higher-order resonances, particularly the 121:118 resonance. This is clearly evident in the IPS shown in Figs. 7(d) and 8(d), where the inclusion of the Mimas–Tethys effects has resulted in the loss of the resonant domain (separatrices), giving rise to several periods of no physical significance. The combined analysis of these dynamic maps and the IPS was crucial in achieving this interpretation.

Section 3.2 addressed the secular dynamics of the Prometheus–Pandora system and the influence of the Saturn term $J_2$ on their orbital evolution. Our analysis revealed the absence of forced components in the eccentricity and inclination of the satellite (see Fig. 11). Nevertheless, the term $J_2$ is identified as a key contributor to the generation of finite amplitudes in these elements, as illustrated in Figs. 9(c), (e), 10(c) and (e). The approximately 6.2-year period observed in the evolution of $\Delta\varpi$ corresponds to the secular timescale over which configurations of pericentre alignment and anti-alignment between Prometheus and Pandora recur. Although the system exhibits circulation of the pericentre difference rather than the pericentre itself, this secular evolution plays a dynamically significant role. In particular, it periodically modulates the relative orbital geometry, giving rise to a natural phase-protection mechanism. However, their secular periods are not identical; the slight differences observed in their pericentre and nodal spectra arise from distinct precession rates imposed by Saturn's oblate gravitational field, primarily due to their different semi-major axis, and are further shaped by asymmetric mutual perturbations and inclination-related effects. By regulating the timing of mutual approaches, this secular cycle reduces the likelihood of close encounters and potential collisions, thus enhancing the long-term stability of the system (see Fig. 12(b)).

Based on the analyses presented, the application of the Continuous Wavelet Transform (CWT) proved to be a robust tool for investigating the time evolution of independent frequencies in the Saturnian moon system. In the simplified three-body scenario (Saturn–Prometheus–Pandora), the dynamics of Prometheus and Pandora's frequencies are predominantly regular, exhibiting long-period oscillations and amplitude modulation. The calculated diffusion coefficients, such as $\delta f_i \approx 10^{-4}$ near nominal positions, are consistent with the periodic motion. Even in resonant regions, the $\delta f_i$ values remain low, indicating dynamic control by regular structures.

However, the inclusion of perturbations from the Mimas–Tethys pair substantially alters the dynamical scenario. The original resonant structure is destroyed, and the frequency behaviour becomes more complex and differentiated for each satellite. While for Prometheus, $\delta f_\varpi$ exhibits coupling to orbital motion and gradual diffusion, with $\delta f_\Omega$ driving chaotic diffusion in certain regions, for Pandora, frequencies show short-period oscillations and large-amplitude irregular fluctuations, with $\delta f_\Omega$ dominating chaotic diffusion near its nominal position. Despite the significant increase (up to 100 times) in diffusion coefficients in some phase space areas, the global values of $\delta f_i$ remain below $10^{-2}$, suggesting that, although the dynamics are more complex and diffusive, they do not reach the threshold for strong chaos.

Combining the mapping of the orbital neighbourhoods of Prometheus and Pandora with their respective IPS (see Figs. 14–16) and considering the perturbative effects of the Janus–Epimetheus pair enabled identification of the domains of the $\Sigma$ multiplets associated with the 17:15 (Epimetheus–Prometheus) and 21:19 (Epimetheus–Pandora) resonances. Examining this phase space revealed that these resonances cause minor alterations to the orbits of Prometheus and Pandora, which do not impact the stability of the Prometheus–Pandora system.

Finally, Section 6 details the analysis of Pandora's tidal evolution, revealing that its past orbital expansion was driven by Saturn's dissipative forces. This mechanism probably resulted in the intersection of several resonances, which suggests the formation of a hypothetical co-orbital system (see Fig. 19(a)). However, the narrow width of these resonances (see Table 6) prevented immediate and stable capture. Consequently, continuous tidal migration guided the Prometheus–Pandora system to the 121:118 resonance, establishing Pandora's current orbit and shaping the observed system dynamics.

A comprehensive analysis of the dynamics of Prometheus and Pandora revelled the intricate nature of their interactions and the remarkable robustness of the system. We demonstrated that the location of both satellites within the separatrices of the 121:118 resonance leads to irregular time evolution of their critical angles. This is evidenced by significant perturbations in their orbital elements and chaotic motions in the IPS. Although the effects of perturbation by Mimas and Tethys are important for the deformation and disappearance of higher-order resonant structures and separatrices, as observed in dynamic maps and IPS, the system remains stable under this influence, being governed by weak chaos. Our studies also revealed the absence of forced terms in the eccentricity and inclination of both satellites, with the term $J_2$ contributing to the generation of amplitudes in these elements. Although we considered the Janus–Epimetheus pair, our results suggest that they do not significantly impact the orbits of Prometheus and Pandora. Finally, our analysis of Pandora's tidal expansion revealed that the satellite crossed several first-order resonances in the past. However, the probability of capture was low.

## CRediT authorship contribution statement

**Demétrio Tadeu Ceccatto:** Methodology, Investigation, Formal analysis, Conceptualization. **Gabriel Teixeira Guimarães:** Writing – review & editing, Methodology, Formal analysis. **Nelson Callegari Jr.:** Supervision, Formal analysis.

## Declaration of generative AI and AI-assisted technologies in the manuscript preparation process

During the preparation of this manuscript, the authors used generative AI tools, including ChatGPT (OpenAI) and Gemini (Google), as well as the AI-assisted translation tool DeepL Write, to support language editing, improve clarity, and refine academic writing of certain sections. After using these tools, the authors carefully reviewed, edited, and validated all content and assume full responsibility for the scientific accuracy, originality, and integrity of the published article.

## Declaration of competing interest

The authors declare the following financial interests/personal relationships which may be considered as potential competing interests: Nelson Callegari Jr reports financial support was provided by São Paulo Research Foundation (FAPESP). If there are other authors, they declare that they have no known competing financial interests or personal relationships that could have appeared to influence the work reported in this paper.

## Acknowledgements

The authors gratefully acknowledge the financial support provided by the São Paulo Research Foundation (FAPESP) through grant no. 2020/06807-7. We also thank Karina Gimenes for her valuable contribution to this work, particularly for carrying out the calculations of the tidal evolution of Pandora's semi-major axis and eccentricity. We also wish to thank the anonymous reviewer for their careful reading of the manuscript and for their valuable suggestions, which contributed to improving the clarity and quality of the text.

## Appendix. Pandora's proximity to the 3:2 resonance with Mimas and its resulting orbital effects

Mimas plays a significant role in generating resonant and secular disturbances on several of Saturn's small satellites. In this context, Pandora is positioned close to the 3:2 resonance with Mimas. Fig. A.21 shows a mapping similar to Fig. 1(b), but considers Pandora clones under the perturbative effects of Mimas and Tethys on Pandora clones, while ignoring the effects of Prometheus. Several diagonal structures corresponding to higher-order resonances are observed in this map, while the 3:2 resonance appears as a wide vertical band in bluish-white tones (associated with small **N** values). The small value of RA = 1% suggests minimal perturbations of the Pandora orbit.

A Pandora clone located in an orbit inside the 3:2 resonance (yellow circle in Fig. A.21), with the same orbital elements given in Table 2, but with a osculating semi-major axis of 142,395 km (equivalent to $\sim$ 141,731 km, in geometric), the oscillation of the geometric angle $\phi_2^{3:2} = 3\lambda_M - 2\lambda_S - \varpi_M$, where $M$ and $S$ refer to Mimas and the of Pandora, respectively, is shown in Fig. A.22(b).

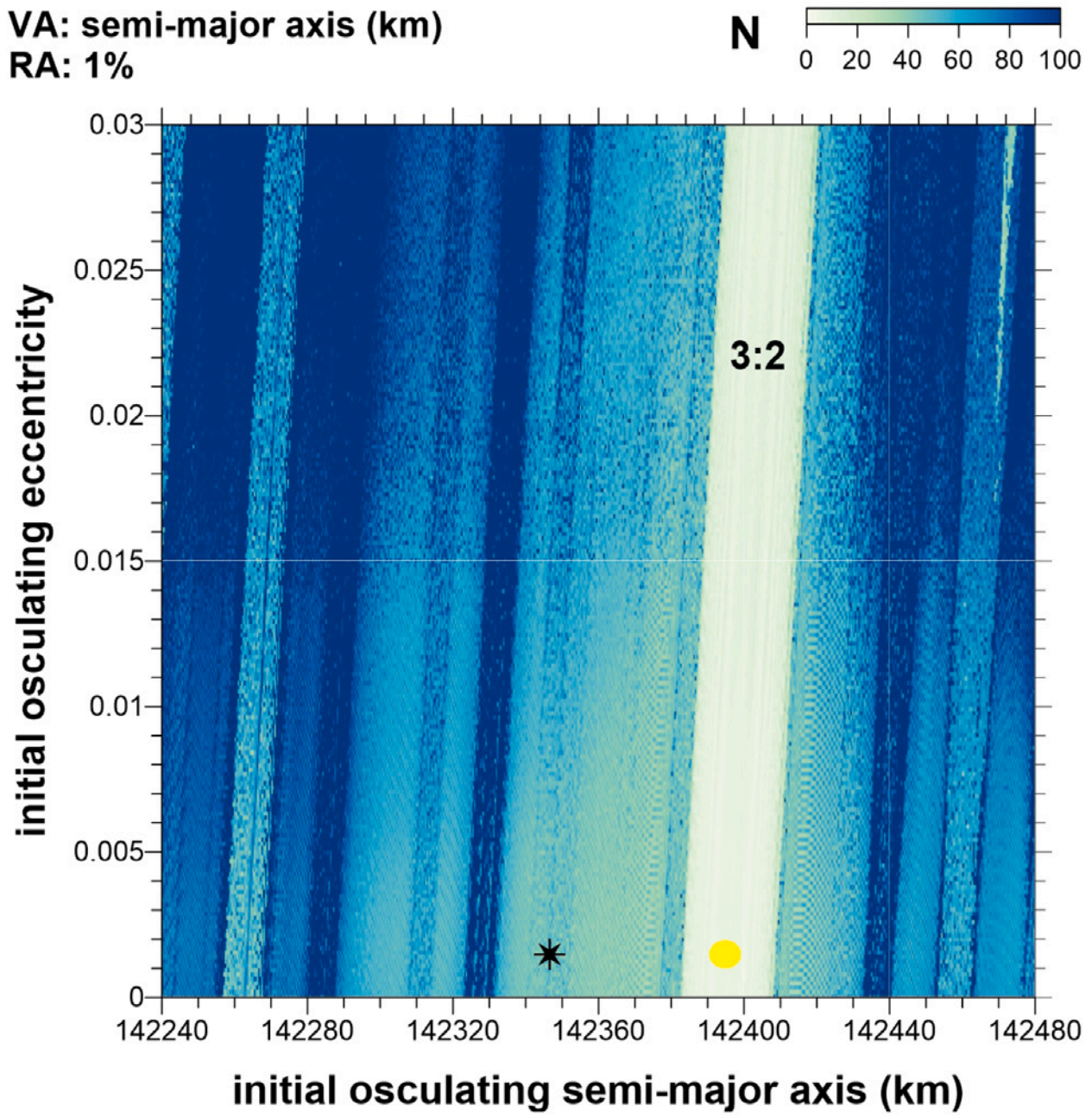

**Fig. A.21.** Similar to Fig. 1(b), but it incorporates the perturbative effects of Mimas and Tethys on Pandora's orbit, without considering the perturbative effects of Prometheus. This dynamical mapping illustrates the region corresponding to the 3:2 resonance between Mimas and Pandora (represented by the bluish diagonal band). The black star shows Pandora's current orbital position, which is approximately 49 km to the left of this resonance. The yellow circle represents a clone of Pandora with an osculating semi-major axis of $a_0$ = 142.395 km and an osculating eccentricity of $e_0$ = 0.0015.

The behaviour of the geometric semi-major axis of this Pandora clone, as shown in Fig. A.22(a), reveals a regular oscillatory behaviour that is distinct from that observed in Fig. 9(b). This enables us to determine the resonant period, $P_{\phi_2^{3:2}} \sim 4301.85$ days, and the period associated with the 4:2 Tethys–Mimas resonance (Greenberg, 1973), $P_{4:2} \sim 38,716.65$ days. Based on Pandora's orbital position (indicated by the black star in Fig. A.21), we can see that Pandora is approximately 49 km behind this resonance.

To investigate the influence of the Mimas–Tethys system on the orbit of Pandora, we performed a spectral analysis in two different scenarios, fixing the eccentricity at $e_0$ = 0.0015: (a) Pandora with the addition of the Mimas–Tethys pair (excluding Prometheus disturbances); and (b) as in (a), but with a Pandora clone considered located within the 3:2 resonance (yellow circle in Fig. A.21). Fig. A.23 shows the spectra of Pandora's semi-major axis, obtained using the FFT algorithm. In scenario (a), the oscillations observed in Fig. 9(d) are eliminated, leaving a principal period of $P \sim 1219.27$ days, with a small amplitude of $\sim$ 0.76 km (see the detailed view in Fig. A.23(a)). In scenario (b), the spectrum shows a small number of peaks, characteristic of regular orbits within a resonance (as observed in Callegari and Yokoyama (2020)), with a period of $P_{\phi_2^{3:2}} \sim 4301.85$ days associated with the resonance angle $\phi_2^{3:2}$, producing an amplitude of 4.73 km in the semi-major axis of this clone. Furthermore, the period $P_{4:2} \sim 38,716.65$ days is related to the 4:2 resonance between Mimas and Tethys. We also identified the harmonic $H_1 \sim 12,905.55$ days, corresponding to $\sim \frac{1}{3}P_{4:2}$. In this spectrum, the 3:2 resonance contributes to the stability of the Pandora clone's orbit (see Fig. A.21).

Fig. A.24 shows the IPS for a neighbourhood of the Pandora clone located within the 3:2 resonance. Analysing the IPS for the semi-major axis (see Fig. A.24(a)), it can be seen that the resonance domain is

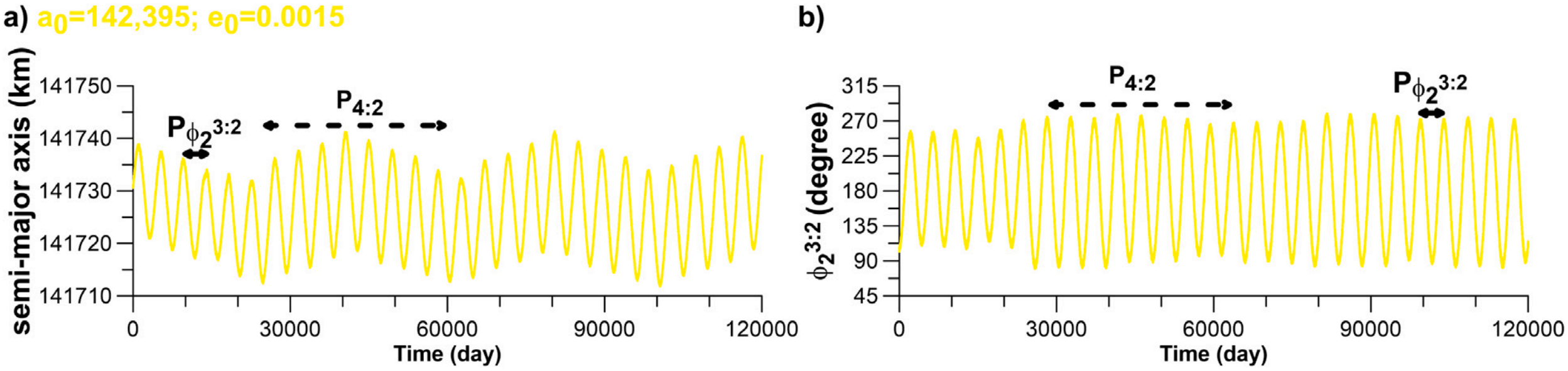

**Fig. A.22.** Temporal variation of (a) the geometric semi-major axis and (b) the geometric angle $\phi_2^{3:2}$ associated with the 3:2 resonance with Mimas, is shown for the Pandora clone, represented by the yellow circle in Fig. A.21. The period $P\phi_2^{3:2}$ shows a variation of 4.73 km, while the period associated with the 4:2 Tethys–Mimas resonance, $P_{4:2}$, generates an amplitude of 2.09 km in the semi-major axis of this clone.

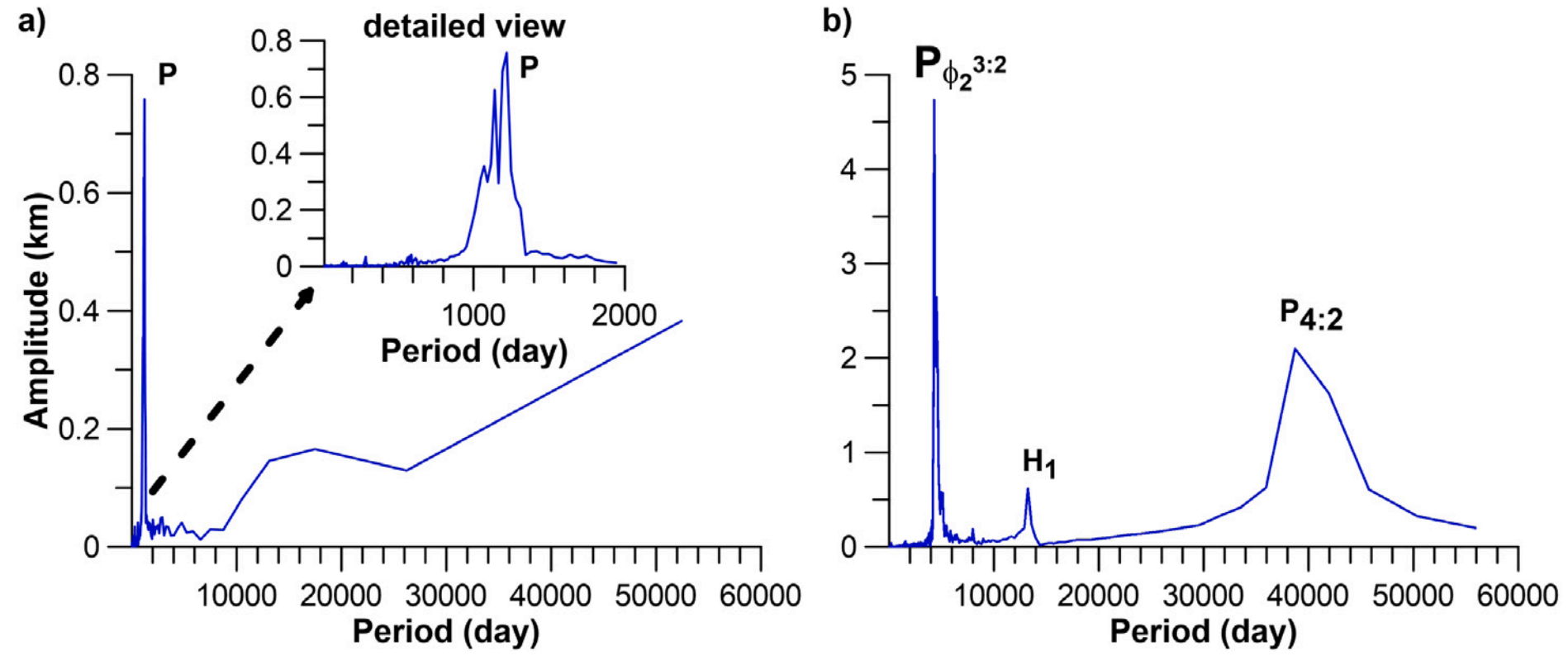

**Fig. A.23.** FFT analysis of the Pandora semi-major axis in two different scenarios: (a) including the perturbative effects of the Mimas–Tethys pair (without Prometheus disturbances); and (b) as in (a), but with a Pandora clone within the 3:2 resonance considered. In scenario (a), including of the Mimas–Tethys pair eliminated the periods shorter than 10,000 days observed in Fig. 8(a), revealing a dominant period of 1219.27 days with an approximate amplitude of 0.76 km. In scenario (b), when analysing the orbit within the 3:2 resonance, the spectrum is characteristic of a regular orbit, with few peaks, particularly the resonant period $P_{\phi_2^{3:2}} \sim 4301.85$ days and the period $P_{4:2} \sim 38,716.65$ days, which are associated with the 4:2 Tethys–Mimas resonance. These produce semi-major axis amplitudes of 4.73 km and 2.09 km for the Pandora clone within the 3:2 resonance, respectively.

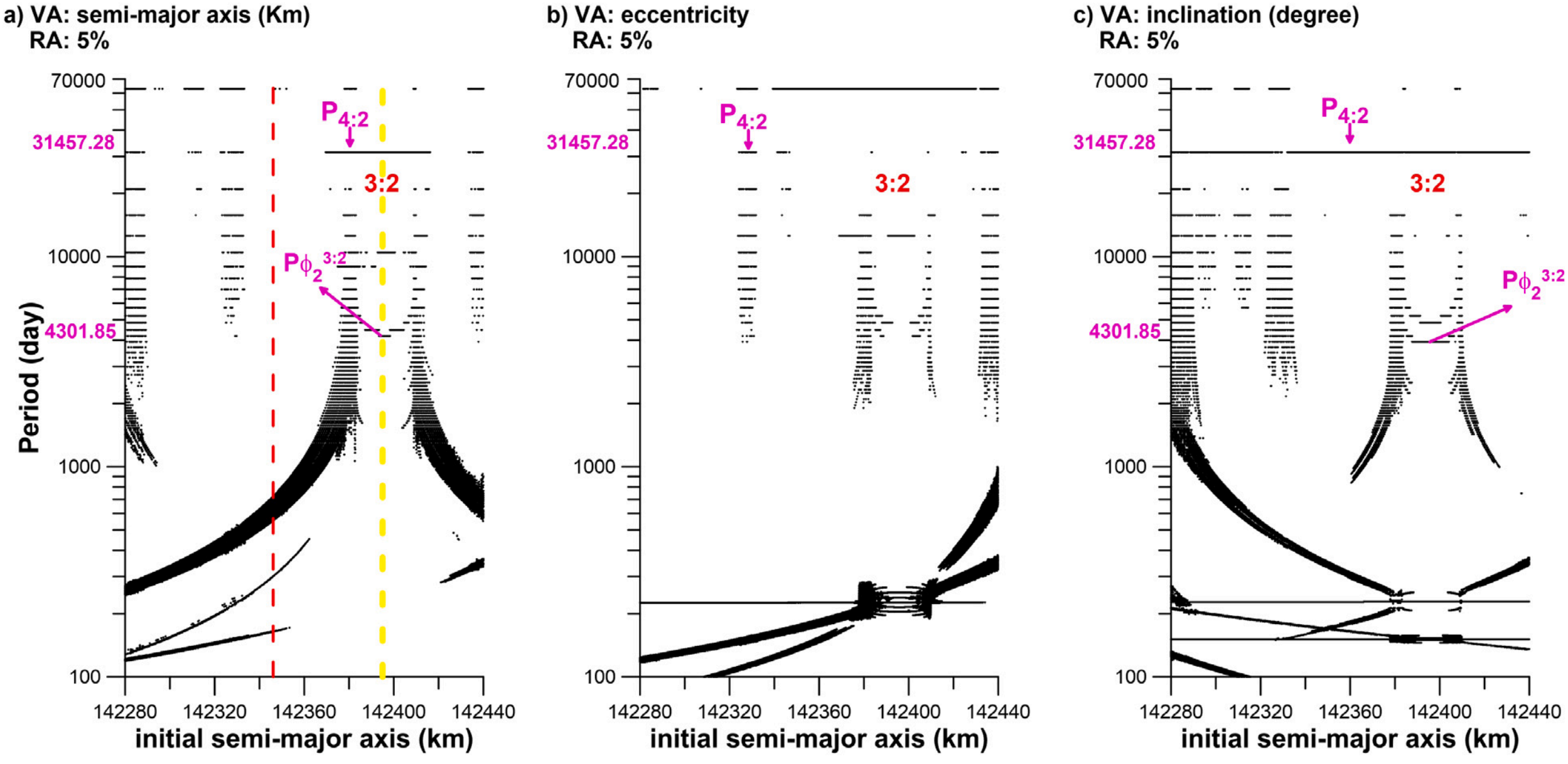

**Fig. A.24.** IPS for the Pandora clone is located inside the 3:2 resonance. The vertical dashed lines indicate the current semi-major axis of Pandora (red) and that of the analysed clone (yellow). (a) The 3:2 resonance domain is clearly defined by its separatrices and shows periods associated with the angle $\phi_2^{3:2}$ ($P_{\phi_2^{3:2}} \sim 4301.85$ days) and the 4:2 Tethys–Mimas resonance ($P_{4:2} \sim 38,716.65$ days). (b) Only periods longer than 200 days, linked to the separatrices, are visible, and the 4:2 resonance has negligible influence on the eccentricity. (c) Both periods are clearly present in the inclination.

clearly defined by its separatrices, with the internal region exhibiting few periods. This includes the periods associated with the angle $\phi_2^{3:2}$ the Tethys–Mimas resonance $P_{4:2}$ seen in Fig. A.23(b). In contrast, Fig. A.24(b), which shows the eccentricity IPS, reveals that the 3:2 resonance domain is less evident, with only periods of more than 200 days visible. These periods are related to separatrices. It can also be seen that the 4:2 resonance has no significant influence on the clones' eccentricity within the 3:2 resonance domain. Finally, Fig. A.24(c), which refers to the orbital inclination, clearly reveals both periods: the one associated with the angle $\phi_2^{3:2}$ and the $P_{4:2}$ resonance.

## Data availability

Data will be made available on request.